\documentclass[a4paper,twocolumn,10pt,noarxiv]{quantumarticle}

\usepackage[numbers,sort&compress]{natbib}

\usepackage[T1]{fontenc}
\usepackage{amsmath,amssymb,amsthm,mathtools}
\usepackage{graphicx}
\usepackage{tikz}
\usepackage{booktabs}
\usepackage{hyperref}
\usepackage{xurl}
\usepackage{cleveref}
\usepackage{enumitem}
\usepackage[normalem]{ulem}
\usepackage{bm}
\usepackage{xspace}

\usepackage{physics}

\crefname{remark}{Remark}{Remarks}
\Crefname{remark}{Remark}{Remarks}
\crefname{proposition}{Proposition}{Propositions}
\Crefname{proposition}{Proposition}{Propositions}

\newcommand{\errobars}{Error bars span the values whose likelihood is within a factor of 100 of the maximum ($\sim3\sigma$) in likelihood relative to the maximum-likelihood estimate, see Appendix \ref{app:error-analysis}.}
\newcommand{\RotThFi}{\textbf{\sffamily Rot$\bm{_{3/5}}$}\xspace}
\newcommand{\RotTh}{\textbf{\sffamily Rot$\bm{_{3}}$}\xspace}
\newcommand{\RotFi}{\textbf{\sffamily Rot$\bm{_{5}}$}\xspace}
\newcommand{\RotThFo}{\textbf{\sffamily Rot$\bm{_{3\times 4}}$}\xspace}
\newcommand{\RegThFi}{\textbf{\sffamily Reg$\bm{_{3/5}}$}\xspace}
\newcommand{\RegTh }{\textbf{\sffamily Reg$\bm{_{3}}$}\xspace}
\newcommand{\RegFi}{\textbf{\sffamily Reg$\bm{_{5}}$}\xspace}

\title{Comparing magic state cultivation methods using matrix
product states}

\author{Tom Hartweg}
\affiliation{QPerfect SAS, European Center for Quantum Sciences, 23 rue du Loess, Strasbourg, 67200, France}
\affiliation{University of Strasbourg and CNRS, CESQ and ISIS (UMR 7006)}
\author{Asier Pi\~neiro Orioli}
\affiliation{QPerfect SAS, European Center for Quantum Sciences, 23 rue du Loess, Strasbourg, 67200, France}

\begin{document}

\begin{abstract}
Magic state cultivation prepares high-fidelity magic states at low expected space-time costs; however, the exact performance of some schemes is unsettled due to the difficulty in simulating non-Clifford circuits.
Here, we use matrix-product states (MPS) based methods to compute the exact performance of two types of fold-transversal cultivation schemes: (i) the Sahay \emph{et al}~\cite{sahayFoldtransversalSurfaceCode2025} method based on the regular surface code S gate, and (ii) a method we propose based on a partially fault-tolerant fold-transversal S gate.
We show that for the former protocol at $d=5$, the $\ket{T}$ output reaches similar logical error rates to the $\ket{S}$ output, traditionally used as a cheap full Clifford proxy. This contrasts with the $\sim10\times$ discrepancy reported for the
$d=5$ colour-code scheme of Gidney \emph{et al.} We also find that our new $d=5$ scheme has $\sim1.3\times$ lower
expected space-time cost while still reaching $10^{-9}$ logical error rate.
We show that MPS and Clifford-augmented MPS (CAMPS) perform on par with or even better than the recently introduced near-Clifford simulator Clifft on the hardest $d=5$ regular surface code scheme.
Additionally, to speed up simulation, we propose a new pre-screening method based on simple Pauli propagation, lowering by up to three orders of magnitude the required number of exact simulations, and use several simulator-agnostic sampling methods such as subset sampling.
\end{abstract}

\section{Introduction}

The main challenge in fault-tolerant quantum computation is finding efficient schemes to implement logical qubits and gates with high fidelity and low spacetime overheads.
Efficient Clifford simulators like Stim \cite{gidneyStimFastStabilizer2021} have led to tremendous progress in fault-tolerant primitives and architecture design through classical simulation, as they allow competing schemes to be compared quantitatively ahead of their experimental realisation.
However, non-Clifford circuits are more challenging to simulate.

An important set of protocols to evaluate are those used to prepare high-fidelity magic states, for example $\ket{T}=T\ket{+}\propto\ket{0}+e^{i\pi/4}\ket{1}$, which allow the implementation of fault-tolerant non-Clifford gates~\cite{gottesmanDemonstratingViabilityUniversal1999,zhouMethodologyQuantumLogic2000}.
Magic state distillation~\cite{bravyiUniversalQuantumComputation2005} represents a long-standing solution with known error suppression, but it comes at a high resource cost.
Recent magic state cultivation (MSC) methods~\cite{gidneyMagicStateCultivation2024,sahayFoldtransversalSurfaceCode2025,Vaknin_2025,chenEfficientMagicState2025, claesCultivatingStatesSurface2025, hiranoEfficientMagicState2025} promise to lower space-time costs compared to pure distillation and thus bring applications closer to reality~\cite{websterPinnacleArchitectureReducing2026a, cainShorsAlgorithmPossible2026}; however, simulation is challenging due to their complex non-Clifford nature and low error rates, making their exact performance harder to estimate.

\begin{figure*}
    \centering
    \includegraphics[width=0.49\textwidth]{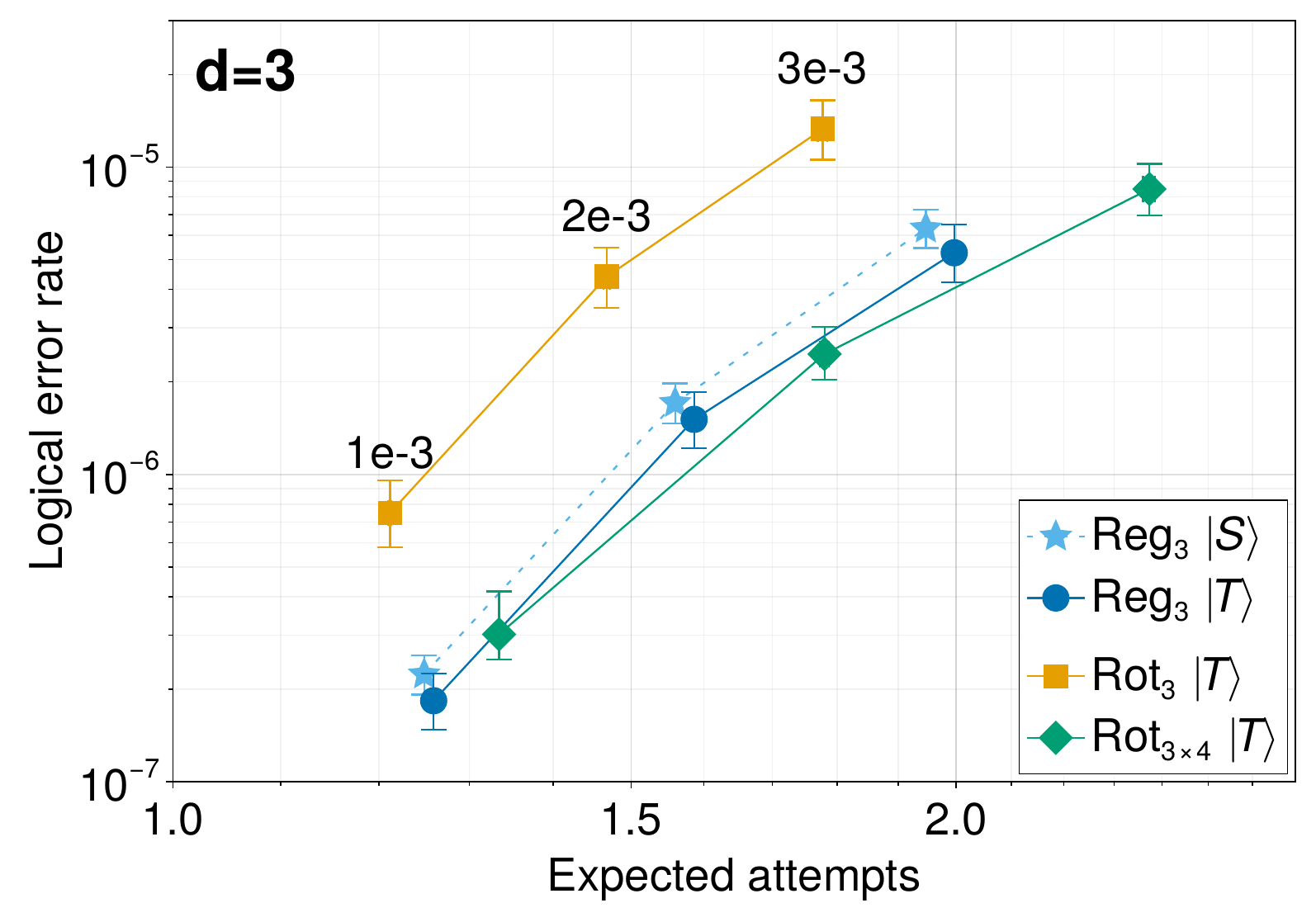}
    \includegraphics[width=0.49\textwidth]{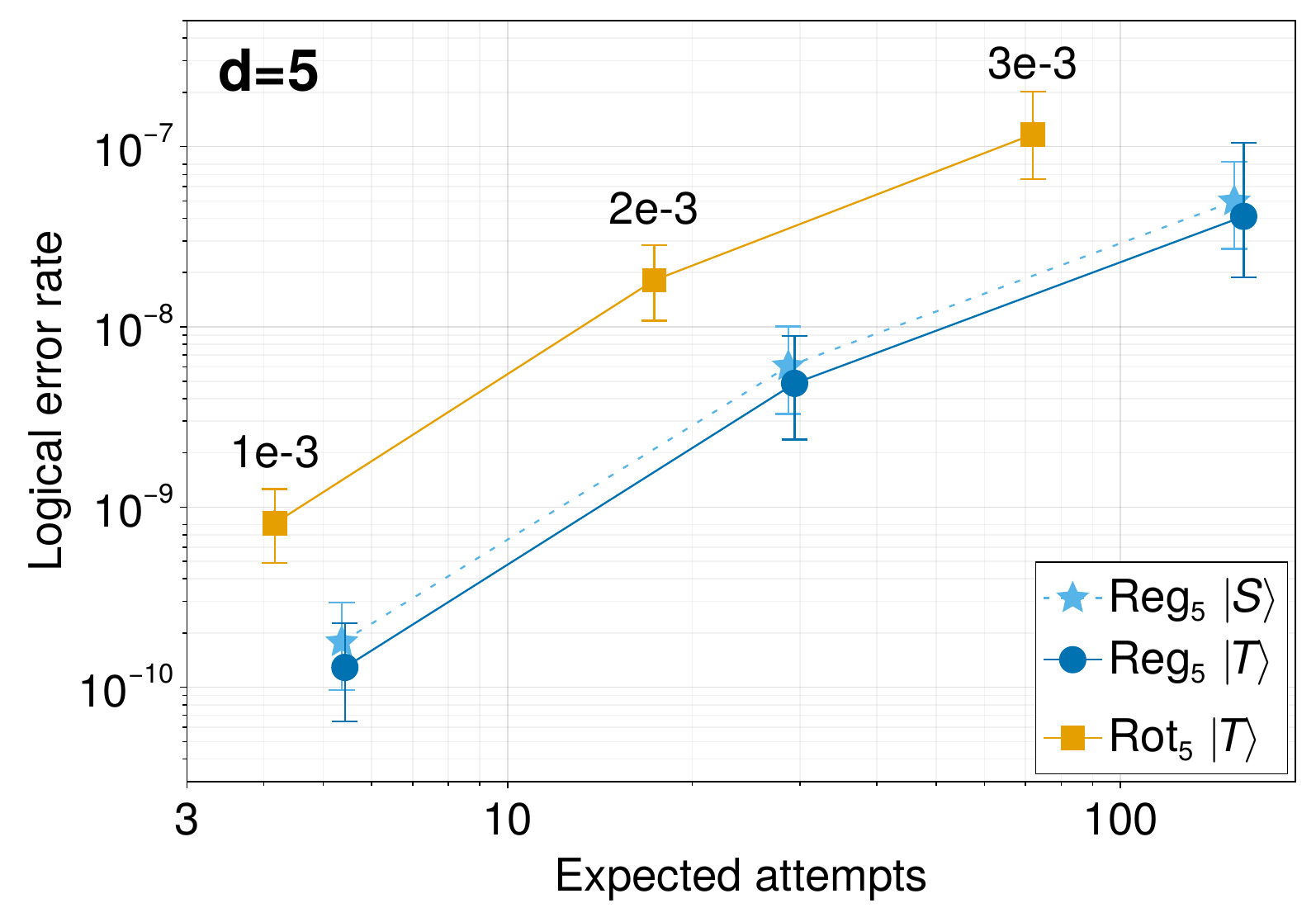}
    \caption{\textbf{Performance of cultivation schemes under uniform depolarising noise} for $\ket{S}$ and
    $\ket{T}$, at output code distance $d=3$ and $5$. \RegThFi are based on Sahay \emph{et al.} \cite{sahayFoldtransversalSurfaceCode2025} circuits, which make use of the regular surface code fold-transversal S gate. \RotThFi and \RotThFo are circuits proposed in this work, based on the partially fault-tolerant fold-transversal S gate of the rotated surface code. The
    $\ket{S}$ results are obtained with Stim, the $\ket{T}$ results with CAMPS
    simulations. These results are obtained with the addition of the final noiseless syndrome extraction in the post-selection. Except for the $\ket{S}$ results, the three points from each curve are obtained with the same samples through subset sampling and are thus not statistically independent. \errobars}\label{fig:results}
\end{figure*}

Recently, efficient new simulators dealing with non-Clifford circuits have been developed, with methods ranging from stabiliser-rank  \cite{bravyiImprovedClassicalSimulation2016, bravyiSimulationQuantumCircuits2019, surtiEfficientSimulationLogical2026, kissingerSimulatingQuantumCircuits2022, sutcliffeFastClassicalSimulation2025, 
wanSimulatingMagicState2026, haenelTsimFastUniversal2026} to generalized tableau~\cite{yoderGeneralizationStabilizerFormalism,liSOFTHighperformanceSimulator2025, chaseClifftFastExact2026, fangSymFTUniversalFaultTolerant2026, tuloupComputingLogicalError2026} or Pauli propagation~\cite{rall2019paulipropagation,rudolph2026paulipropagation, queraPPVM2026}. The cost of these methods scales with the stabiliser rank, the active space dimension and the size of propagated Pauli sums, respectively, all of which grow with the non-Clifford content of the circuit.
These advances have enabled the exact simulation of the Gidney \emph{et al.} original colour-code cultivation scheme~\cite{gidneyMagicStateCultivation2024} at $d=5$~\cite{liSOFTHighperformanceSimulator2025, tuloupComputingLogicalError2026}, even end-to-end~\cite{chaseClifftFastExact2026}.
A different line of development explores the use of tensor network-based methods~\cite{ciracMatrixProductStates2021}, which are instead limited by entanglement and quite efficient in simulating universal large-scale quantum circuits~\cite{tindall2024IBMtensornetwork, pan2022sycamoresimulation, tindall2026dwavesim, leonteva2025comparative}.
Specifically, some works have explored the direct use of tensor networks on quantum error correction problems~\cite{Poulin_PRL2017, manabe2025leakageerrors, barone2025colorcode, orioli2026optimizedmps} and others propose extensions to combine them with stabilizer-based methods~\cite{Masot2024stabilizertensor, mello2024hybridstabsmpos, lami2024quantumstatedesignsclifford, qian2024camps, liuClassicalSimulabilityClifford2026}, whose cost scales with Clifford-irreducible entanglement rather than by the total amount of magic.
As shown here, these methods remain useful for cultivation circuits where magic is distributed over many qubits, while the entanglement stays modest enough for tensor network representation.

In this work, we compute the exact performance of cultivation schemes based on fold-transversal implementations of the S gate on the rotated surface code, summarised in Fig.~\ref{fig:results}---such methods are well-suited to rapidly advancing quantum computing platforms such as neutral atoms~\cite{Bluvstein2024, reichardt2025ftqc, yang2026rasql, cainShorsAlgorithmPossible2026, Zhou_2025}.
Specifically, we consider (i) the Sahay \emph{et al} scheme~\cite{sahayFoldtransversalSurfaceCode2025}, based on the fault-tolerant fold-transversal S gate of the regular surface code (\RegThFi), and (ii) a new method based on the \emph{partially} fault-tolerant fold-transversal S gate of the rotated surface code~\cite{holmesQuantumLogicCodes2026} (\RotThFi).
The d=5 version of our protocol reaches a $\sim1.2\times$ lower space-time overhead than the regular surface code scheme while still reaching a logical error rate of $\sim10^{-9}$; a middle ground of interest for near-term applications. Moreover, by simulating the $d=5$ Sahay \emph{et al} cultivation exactly, we show that cultivating $\ket{T}$ states with this protocol gives similar (possibly lower) logical error rates than cultivating $\ket{S}$ states, as was observed for the $d=3$ scheme. This contrasts with the $7\times$ to $30\times$ higher logical error rates observed for the colour code cultivation~\cite{gidneyMagicStateCultivation2024, liSOFTHighperformanceSimulator2025, tuloupComputingLogicalError2026, chaseClifftFastExact2026}, depending on the error rate, which highlights the importance of running exact simulations for each method separately.

To obtain these results, we used matrix product states (MPS)~\cite{ciracMatrixProductStates2021, orioli2026optimizedmps} and Clifford-augmented MPS (CAMPS)~\cite{qian2024camps, liuClassicalSimulabilityClifford2026}. We show that the states along the $d=5$ regular surface code cultivation circuit, the hardest of the whole set, admit a low bond-dimension MPS with noiseless peak bond dimensions $\chi_{peak}=96$, whereas CAMPS reduces it to only $\chi_{peak}=5$.
In comparison, we found that on the same circuit the extended tableau simulators Clifft~\cite{chaseClifftFastExact2026} and SymFT~\cite{fangSymFTUniversalFaultTolerant2026} reach a peak active dimension of $k_{peak}=22$, yielding a factor $2^{12}$ in the dense
active-array cost compared to the colour code cultivation circuit, for which $k_{peak}=10$.
As a result, pure MPS performs only $3\times$ worse than Clifft on the hardest $d=5$ circuit, whereas our CAMPS implementation reaches a $\sim13\times$ improvement in raw\footnote{Without any circuit-specific sampling acceleration.} sampling rate compared to Clifft.

Orthogonally, we introduce a new pre-screening method based on the fact that a large fraction of the measurements of cultivation can be decided by stabiliser propagation alone under Pauli noise despite the presence of non-Clifford gates. Coupled with the post-selection protocol of cultivation, it reduces the number of trajectories needing a full simulation by $\sim 3$ orders of magnitude at a cost of $\sim\mu s$ per trajectory. This method can substantially improve the speed of other MSC simulations under Pauli noise. 
This, combined with CAMPS, reaches a several-kHz sampling rate of the circuit under depolarising noise on a commodity CPU. Further leveraging subset
sampling~\cite{tuloupComputingLogicalError2026, heussenDynamicalSubsetSampling2024} allows us to produce the previously mentioned results.

The rest of this paper is organised as follows. In \cref{sec:MSC}, we give an overview of the MSC methods and introduce three new circuits. In \cref{sec:methods}, we go over the simulation methods employed in this work: MPS \ref{sec:mps}, CAMPS \ref{sec:camps}, and Clifft \ref{sec:clifft} used as a comparison. In \cref{sec:simagnostic}, we describe some simulator-agnostic methods helping to reduce or accelerate the required sampling: subset sampling \ref{sec:subset_sampling} and offline \ref{sec:offline_screening} and online \ref{sec:online_screening} screening. Then in \cref{sec:results}, we discuss in detail the results obtained in this work.

\section{Magic State Cultivation}
\label{sec:MSC}

\begin{figure*}
    \centering
    \includegraphics[width=\textwidth]{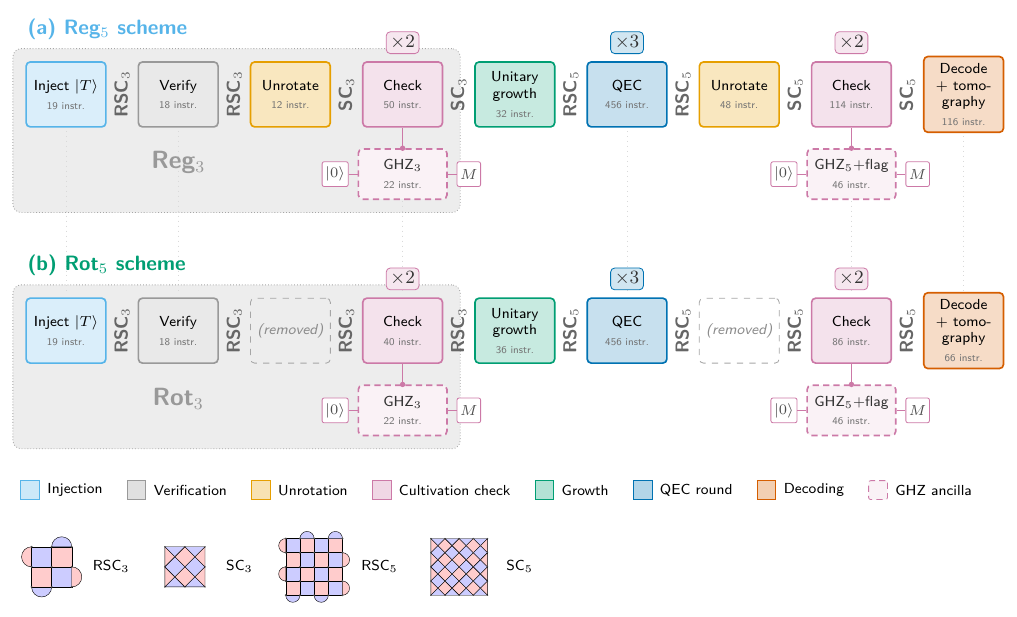}
    \caption{ \textbf{Gadget-level structure of the two $\bm{d=5}$ $\ket{T}$ cultivation
     \RegFi and \RotFi,} both built on the same d=3 rotated surface-code patch
     (17 physical qubits) grown to distance 5 (49 physical qubits). The label between each pair of stages gives the code carried on that
     step -- RSC = rotated surface code, SC = unrotated surface
     code, subscript the code distance. A schematic representation of each code is given at the bottom of the figure. The dotted grey box, labelled
     \RegTh / \RotTh, marks the
     corresponding $d=3$
     circuit prefix. Each cultivation check is a control-$H_{XY}$ gadget driven by a separate GHZ ancilla register --
     3 qubits at $d=3$, 5 qubits $+$ 1 flag qubit at $d=5$ -- prepared and measured
     for each of the two rounds.
     The number of instructions displayed in the boxes is the count over all the repetitions, if applicable.
     \textbf{(a) \RegFi} keeps the patch in the SC
     for both check rounds, which requires an explicit unrotation gadget
     (RSC$\to$SC, 2 layers of CNOT) before each of the two check/growth stages,
     alternating the patch between the RSC and SC. 
     \textbf{(b)} \RotFi performs
     both check rounds directly on the RSC, and
     grows the patch with a single unitary circuit (2 CNOT
     layers), saving the unrotations and requiring fewer operations per check, at the cost of a reduced distance in the Z direction. Both schemes apply 2 rounds of standard
     rotated-surface-code QEC (\texttt{rsc\_qec}) after growth and finish with
     unitary decoding onto a single physical qubit followed by Pauli
     tomography.}\label{fig:circuit_struct}
\end{figure*}

Magic State Cultivation (MSC) consists of three main stages: noisy state preparation, cultivation, and escape (or expansion). The noisy state preparation and cultivation stages are not decodable; thus, only error detection and post-selection are performed to decrease the logical error rate. Because of this, the retry rate increases exponentially with the number of noisy gates and idling qubits, so minimising gate count and depth becomes important. The three main stages involve:

\begin{enumerate}
    \item \emph{Noisy state preparation.} The targeted magic state is prepared in a small distance code using a non-fault-tolerant shallow circuit. This includes methods such as injection or unitary preparation with post-selection. The resulting magic state has a typical logical error rate on the order of the physical error rate.
    \item \emph{Cultivation.} The noisy magic state is then checked using projective measurements of an operator that stabilises it, typically the $H_{XY}$ operator for the $\ket{T}$ state.  After $d-1$ repetitions of the checks, the distance of the initial code becomes a bottleneck for the error detection distance and ``growing'' the code to a larger distance is required. 
    \item  \emph{Escape.} The escape stage fault-tolerantly bridges between the cultivation error detection distance of $d$ and the error correction distance of at least $d$ that is required for the remainder of the computation. This involves growing the size of the code where the $\ket{T}$ is prepared.
\end{enumerate}

In this paper, we will focus on the error rate during the first two stages.
Since the escape stage is exclusively made of Clifford gates, techniques such as hand-off simulations~\cite{sahayFoldtransversalSurfaceCode2025} represent an efficient simulator-agnostic way of evaluating it, provided that the first two parts are simulated exactly.

Sahay \emph{et al.} proposed to perform the noisy state preparation into a $d=3$ rotated surface code and then morph it to a regular $d=3$ surface code, where the fold-transversal logical S gate \cite{moussaTransversalCliffordGates2016} can be used to perform the $H_{XY}=e^{-i\frac{\pi}{4}}SX$ checks.
\Cref{fig:circuit_struct}.a) shows the gadget-level structure of the $d=5$ circuit, and Appendix \ref{app:circuit_description} describes the circuit used in this work in detail.
The $H_{XY}=e^{-i\frac{\pi}{4}}SX$ checks are implemented using a GHZ state-controlled version of the logical gates
(\cite{sahayFoldtransversalSurfaceCode2025}, Fig.~A3/A8). Measuring the GHZ state projects the surface code to the $\ket{T}$ or the $Z\ket{T}$ state. The $H_{XY}$ operator contains $S$ and $CZ$ gates; thus, the controlled version contains $CS$ and $CCZ$ gates, which are non-Clifford. Alternatively, one can cultivate the $\ket{S}$ state using a controlled version of the logical $Y$ gate. This yields a circuit that is purely Clifford and can be efficiently simulated, and thus is traditionally used as a proxy for the $\ket{T}$ state cultivation performance.   

Additionally, we propose and demonstrate a cultivation scheme based on the rotated surface code fold-transversal logical S gate that was recently discovered \cite{holmesQuantumLogicCodes2026}. \Cref{fig:circuit_struct}.b) depicts its gadget-level structure. It reduces the total number of gates in the circuits by removing the morphing circuits between the rotated and regular surface code, thus reducing the retry rate and the per-try space-time overhead. This comes at the cost of a higher logical error rate due to the distance reduction of the new gate in the Z direction (which is why it is partially fault-tolerant). \Cref{fig:results} shows the performance of both the Sahay \emph{et al.} scheme and the proposed one, which we denote by \RegThFi and \RotThFi, respectively, see \cref{sec:comparison} for further discussion.

We also show that the distance reduction of the S gate in the rotated surface code can be mitigated by using an asymmetric rotated surface code of distance 3 and 4 in the X and Z directions, respectively. Indeed, this ensures that the detection distance of the schemes stays at 3 in both X and Z directions. The performance of this scheme, which we call \textbf{Rot$\bm{_{3\times4}}$}, is shown in \Cref{fig:results} and discussed further in \Cref{sec:comparison}. 

\section{Simulation Methods and Optimisations}
\label{sec:methods}

\label{sec:simulators}
\subsection{Matrix Product State}
\label{sec:mps}

A quantum state $\sum_b c_b\ket{b}$ with  $b\in\mathbb{F}_2^N$ can be expressed as a matrix-product state (MPS)~\cite{ciracMatrixProductStates2021} by performing recursive singular value decomposition on $c_b$ (reshaped as a matrix over different bipartitions), leading to:
\begin{equation}
c_{i_1\dots i_N} = \sum_{\alpha_1,\dots,\alpha_{N-1}}
A^{[1]}_{i_1\alpha_1} A^{[2]}_{\alpha_1 i_2 \alpha_2} \cdots
A^{[N]}_{\alpha_{N-1} i_N}.
\end{equation}
Each $A^{[k]}$ is a rank-3 tensor (rank-2 at the boundaries),
with $i_k \in \{0,1\}$ the physical index and $\alpha_k$ the virtual (bond)
index. The bond dimension $\chi_k$ is the number of \emph{nonzero} singular
values at the corresponding SVD step, and can range anywhere from $1$ up to
$2^{\min(k,N-k)}$. $\chi_k$ is directly the
Schmidt rank of $\ket{\psi}$ across the bipartition $\{1,\dots,k\} \,|\,
\{k+1,\dots,N\}$ and the memory footprint of the MPS representation is polynomial in $\chi$.

The tensor chain $A^{[1]}\cdots A^{[N]}$ presupposes a
\emph{linear} ordering of the $N$ qubits. The choice
of ordering is important: it fixes, for every bond $k$, which set of
qubits $\{1,\dots,k\}$ is separated from its complement, and hence which
Schmidt rank $\chi_k$ the corresponding bond must support. A poor ordering
can force $\chi$ far above what the circuit's entanglement structure
actually requires; this is the rainbow problem mentioned in \cite{orioli2026optimizedmps}.
Moreover, the cost of applying multi-qubit gates to an MPS grows with the distance in the 1D chain between the involved qubits, as the operation needs to touch all bonds sitting in between. 

This is why careful optimisation of the qubit order is needed to achieve optimal computational performance. Cultivation is designed around a 2D structure, the surface code, making it challenging to embed the 1D chain for the MPS.
However, we can use the stabiliser formalism to optimise the qubit order: the bipartite entanglement of a stabiliser state grows with the minimal number of shared stabilisers between the two parts~\cite{fattal2004entanglement}, and so does the bond dimension required. While our circuits are not completely Clifford, this is still a useful heuristic.

\begin{figure}
    \centering
    \includegraphics[width=\linewidth]{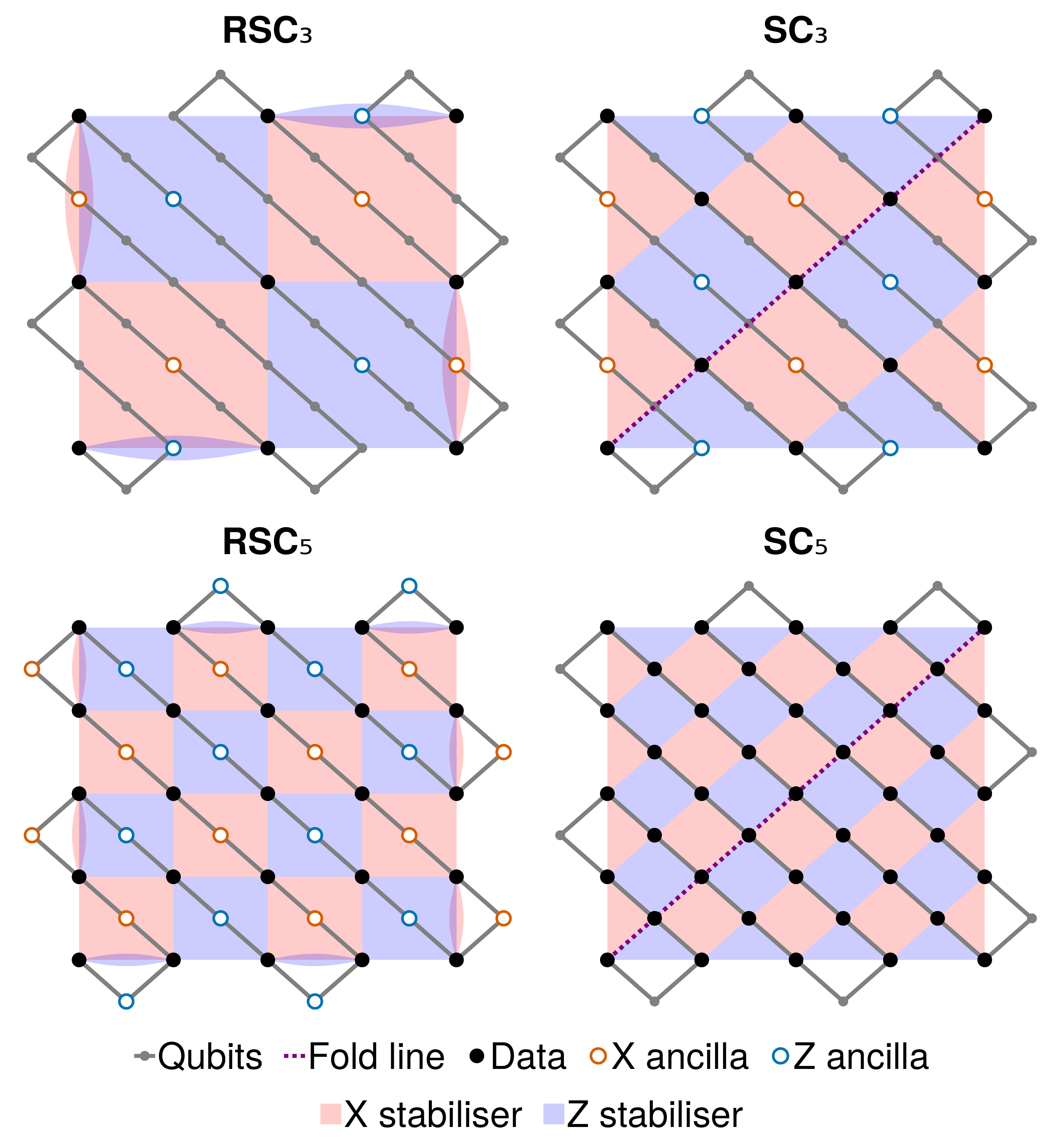}
    \caption{\textbf{Qubit layout for the \RegFi cultivation MPS simulation.} The distance 3 rotated surface code (RSC$_3$) is used at the injection stage, the distance 3 regular surface code (SC$_3$) is used during the d=3 $H_{XY}$ operator checks, the distance 5 rotated surface code (RSC$_5$) is used for the intermediate syndrome extraction, and the distance 5 regular surface code (SC$_5$) is used to do the d=5 $H_{XY}$ operator checks. The ancillas of the SC$_3$ are only used to perform the rotation to the RSC$_5$ code.}
    \label{fig:layout_d5}
\end{figure}

The proposed qubit layout, described in \Cref{fig:layout_d5} for \RegFi, aims to reduce the number of shared stabilisers across cuts at each step of the cultivation process. The stabilisers change from rotated and regular surface codes and from distance 3 to 5 throughout the circuit, as shown in the figures. The embedding of the distance 3 codes inside the distance 5 codes reduces long-range gates during the transformation between the two, increases qubit reuse, and still gives a good ordering for the distance 3 code. Finally, the layout orientation is chosen to reduce the range of the multi-qubit gates during the $H_{XY}$ checks, which create connections between the qubits sitting across the fold-symmetry line of the regular surface code used to implement the S gate (the dashed line on the figures). Additional details, as well as the layouts used for the other circuits, are given in Appendix~\ref{app:qubit_layout}.

\subsection{Clifford augmented MPS}
\label{sec:camps}

In this work, we also make use of the Clifford augmented MPS (CAMPS) representation and circuit simulation technique~\cite{qian2024camps}, following the scheme of Liu and Clark~\cite{liuClassicalSimulabilityClifford2026}. We give here only a brief description of the method. CAMPS allows one to decouple Clifford entanglement from genuine non-Clifford entanglement by factoring the state $\ket{\psi}$ into a Clifford circuit $C$, called a Clifford frame, and a residual state $\ket{m}$, called the magic residual or Clifford coordinates, such that $\ket{ \psi} = C\ket{m}$. This representation shows a gauge redundancy, as one can insert an identity in the form $\tilde{C}\tilde{C}^\dagger$, with $\tilde{C}$ an arbitrary Clifford circuit, and absorb one part into the frame and one part into the magic residual:
$$
\ket{\psi} = C\ket{m} = C\tilde{C}\tilde{C}^\dagger\ket{m}=C'\ket{m'}
$$
Here, $C'=C\tilde{C}$ and $\ket{m'}=\tilde{C}^\dagger\ket{m}$. The Clifford frame can be efficiently represented by a tableau, so the gauge redundancy can be used to search for a gauge in which the magic residual exhibits a desired property.
CAMPS proposes to use an MPS representation for the magic residual and to use the gauge freedom to find low bipartite entanglement representations; this process is called \emph{disentangling}. This procedure is only able to remove Clifford-type entanglement \cite{fuxDisentanglingMagicStates2025}, so the cost of this simulation method depends on the amount of genuine non-Clifford entanglement.

During simulation, Clifford operations can simply be added in the Clifford frame, only costing the tableau update and never raising the bond dimension of the magic residual MPS. Non-Clifford gates and projectors are represented as weighted sums of Pauli strings, propagated through the tableau and applied to the magic residual, potentially increasing the bond dimension.
A disentangling procedure is then usually performed after such an operation. For near-Clifford circuits, in particular quantum error correction circuits, most of the entanglement comes from the Clifford gates. Thus, CAMPS remains efficient. 

We explored the sensitivity of CAMPS to the 1D qubit ordering of the simulated circuit. As the SWAP gate is a Clifford operation, a particular 1D layout of the qubits can be described as a SWAP circuit and absorbed by the Clifford frame. Thus, we would expect the disentangler to be able to find the optimal layout to reduce the entanglement in the magic residual. However, the state-of-the-art disentangler is a greedy local search, sweeping along the 1D layout to find entanglement-lowering two-qubit Clifford gates acting on nearest neighbours only. Thus, it will not necessarily find the optimal.

We found that starting the CAMPS simulation with the qubit layout presented in the previous section reduced the simulation time of CAMPS by $25\%$ on average compared to random ordering, showing significantly less dependence than pure MPS on qubit ordering. All simulations presented below are performed with the same layouts as pure MPS. Further exploration of this dependency is left for future work.

\subsection{Clifft}
\label{sec:clifft}

To compare MPS and CAMPS against state-of-the-art methods, we also run simulations with Clifft \cite{chaseClifftFastExact2026}. Clifft is similar in spirit to CAMPS in the sense that it also separates the state into a Clifford frame and a magic residual; however, the magic residual is represented using a state vector.
State vectors allow to apply rapid arbitrary operations at the cost of storing $2^N$ coefficients.
Instead of removing entanglement from the magic residual as CAMPS, Clifft uses the gauge choice of the Clifford frame representation to reduce the number of qubits on which genuine operations are performed.
By restricting the state-vector representation to those qubits, the effective $N$ is small, and the exponential scaling is mitigated.
This allows one to reduce the 44-qubit $d=5$ cultivation circuit of Gidney \emph{et al.} to a peak active dimension $k_{peak}$ --- the maximum number of qubits reached in the magic residual state vector --- of 10, sampling the circuit at almost $MHz$.

We did not find any dependence on the circuit layout for the runtime of Clifft; thus, we run all the simulations described below with the same layout as MPS and CAMPS.

\section{Efficient sampling methods}
\label{sec:simagnostic}

We discuss here several simulator-agnostic methods to reduce the amount of noise sampling required.
We consider the cultivation circuits in the presence of uniform depolarising noise with probability $p$, i.e.~any of the $z=4^k-1$ $k$-qubit non-identity Pauli strings has probability $p/z$.
A given shot samples one particular Pauli for each noise channel, producing a fault configuration $\mathcal{F}$. The number of channels that resolved to a non-identity Pauli string is called the weight of the fault configuration $w(\mathcal{F})$. 
After simulation, a shot is kept if every measurement outcome is zero and discarded otherwise, with nothing decoded and no correction applied.
We flag a logical error whenever the end logical state overlap with a $\ket{T}$ state is not perfectly 1, up to machine precision. Each shot therefore ends in one of three outcomes: discarded, accepted and correct, or accepted and
logically wrong.
We report the acceptance rate $p_{acc}$
together with the logical error rate given acceptance, $p_{le}$.

\subsection{Subset sampling}
\label{sec:subset_sampling}

Drawing configurations directly from the probability distribution of the noise
channels is prohibitive. Due to the fault-tolerant nature of the circuit, failures
are a rare outcome, and as the relative error of the estimate scales as
$1/\sqrt{N p_{acc} p_{le}}$, with $N$ the number of shots, resolving $p_{le}\sim10^{-9}$ demands $N\gg10^{9}/p_{acc}$. However, this is not necessary; the number of faults per shot concentrates around $Mp$, where $M$ is the total number of noise channels, and for realistic
noise strength, $p\sim0.1\%$, almost every shot lands in a low-weight configuration
for which the circuit is designed to never fail, while the high-weight
configurations causing logical errors are drawn least often.

We use subset sampling \cite{bravyiSimulationRareEvents2013a, heussenDynamicalSubsetSampling2024}, following Tuloup and Ayral~\cite{tuloupComputingLogicalError2026} application to cultivation. Subset
sampling rewrites the sampling
distribution as a sum of terms with different \emph{fault weight}
$w$. Since noise channels fire independently and all have the same probability $p$
of sampling a non-identity Pauli, the probability of a shot having weight $w$ is an
exactly known binomial factor
$\mathrm{Bin}_w(p)=\binom{M}{w}p^{w}(1-p)^{M-w}$. Thus
\begin{equation}\label{eq:reweight}
\begin{split}
    p_{acc} &= \sum_{w}\mathrm{Bin}_w(p)\,p^{(w)}_{acc}\\
    p_{acc}\,p_{le} &= \sum_{w}\mathrm{Bin}_w(p)\,p^{(w)}_{acc}p^{(w)}_{le}
\end{split}
\end{equation}
with $p_{acc}p_{le}$ the joint accept-and-fail probability. The
$p^{(w)}_{acc}, p^{(w)}_{le}$ are properties of the circuit at fixed weight and
carry no dependence on $p$; all $p$-dependence sits in the prefactors.

The per-weight errors $p^{(w)}_{acc}$ and $p^{(w)}_{le}$ can be estimated by grouping the shots with the same fault weight. Out of $n_w$ such shots, $n_w^{acc}$
survive post-selection and $n_w^{err}$ of those flip the logical observable, giving
the ordinary binomial proportions
\begin{equation}\label{eq:estimators}
    \hat{p}^{(w)}_{acc}=\frac{n_w^{acc}}{n_w},
    \qquad
    \hat{p}^{(w)}_{le}=\frac{n_w^{err}}{n_w^{acc}},
\end{equation}
which we insert into Eq.~\eqref{eq:reweight} and divide to obtain $p_{le}$.
The gain comes from $n_w$ being free to choose.

Drawing $w$ from $\mathrm{Bin}_w(p)$ itself reproduces the physical-level sampling, and it never costs accuracy relative to direct sampling at equal total shots, up to an $O(1/N)$ correction, improving on it as soon as the conditional rates vary with $w$~\cite[Sec.~5.6]{cochranSamplingTechniques1977} --- which is usually the case for fault-tolerant circuits.
We sample $w$ from $\mathrm{Bin}_w(p)$, at no cost in accuracy compared to direct sampling~\cite[Sec.~5.6]{cochranSamplingTechniques1977}.
In cultivation, $p^{(w)}_{le}=0$ for every $w$ below the circuit error detection distance $d_e$, since every such error configuration will be discarded through post-selection.
So the $p^{(w)}_{le}$ run from exactly zero below $d_e$ to $O(1)$ at high weight. Similarly, $p_{acc}^{(w)}$ is 1 at $w=0$, and then decreasing sharply with increasing $w$. Because of this, we can concentrate our sampling effort on $w\geq 1$ and $w\geq d_e$ for $p^{acc}$ and $p_{le}$, which can save orders of magnitude in the required sample count. 

Importantly, as the largest fault weight possible can be quite high, we need to truncate the sampling at a maximum weight $w_{max}$; the neglected contribution is bounded by $\sum_{w>w_{max}}\mathrm{Bin}_w(p)$. We choose $w_{max}$ so that it sits orders of magnitude below the reported rate. To be certain to reach the $w_{max}$ bins, we spend some of the sampling effort uniformly on the $w$ range $1$ to $w_{max}$.
The estimation of error bars is discussed in App.~\ref{app:error-analysis}.

Finally, because the per-bin rates carry no $p$-dependence, one set of samples can give
the full curve. Importantly, this means that all points evaluated at different values of $p$ are statistically correlated, and must not be treated as independent experiments. Note that, except stated otherwise, all the curves presented in this work are obtained using this technique.

\subsection{Offline screening}
\label{sec:offline_screening}

In \RegFi, 91 out of the 133 measurements can be easily decided without any complete simulation, by simple stabiliser propagation only.
The key is that in the circuits that we are considering, the only non-Clifford gates are diagonal (\textsc{T}, \textsc{CS} and \textsc{CCZ}), so any Pauli string with \textsc{I} or \textsc{Z} on the support of the non-Clifford gates will just commute through.
Starting from $\ket{0}^{\otimes n}$, at each step we track the state stabilisers that commute with the next operator and drop the ones that don't commute.
At every measurement, we check if the Pauli operator associated to the measurement (e.g.~Z for $M_\text{Z}$) is in the group formed by the stabilisers that are still tracked; if yes, then the outcome of the measurement is easily obtained.
Crucially, Pauli noise can only change the sign of the stabilisers (but not the shape) as they propagate through the circuit, which can potentially flip the outcome of the measurement.
Note that only some measurements can be determined in this way.

When using Pauli noise, a pre-screening of the shots can be performed by computing these pre-determinable measurements and discarding the trajectory if any are flipped, thus avoiding the heavy complete simulation entirely.
This method is unbiased, since it cannot create false positives, and can be performed at $\sim\mu s$ scale by precompiling the effect of the noise channels on the measurements, following SymFT \cite{fangSymFTUniversalFaultTolerant2026} GF(2) precompilation. Only the shots for which those 91 measurements are not flipped require a full simulation to decide the 42 remaining measurements. For example, on the $w=5$ strata, $99.7\%$ of the trajectories can be pre-screened in such a way, leaving almost three orders of magnitude fewer samples to simulate. Naturally, this method can be applied to all the cultivation schemes explored here, and is simulator agnostic.

\subsection{Online screening}
\label{sec:online_screening}

For the shots that pass through the offline screening, we allow the full simulation to stop as soon as one measurement returns a flipped value. Interestingly, this method pairs very well with subset sampling and most simulators. Indeed, the higher the fault-weight, the more likely it is for the trajectory to be discarded early.
Moreover, the cultivation circuits grow the distance of the code from 3 at the beginning to 5 towards the end, which is harder to simulate as it contains more active qubits, more entanglement, and more magic.
Thus, being able to stop early saves the most computationally heavy part of the circuit.

\section{Results}
\label{sec:results}

\subsection{Benchmark}
\label{sec:runtime}
In \Cref{tab:runtime_noiseless}, we present the characteristics of all the circuits considered in this work, along with a comparison of the runtime of MPS, CAMPS and Clifft. All the runtimes are produced on the circuits including the uniformly depolarising noise channels, but the noise probability $p$ is set to 0. The MPS runs are performed using the QPerfect state-of-the-art MPS simulator, TensorWeaver, with the 1D qubit layout described in \cref{sec:mps}. The CAMPS runs are performed on an internal implementation of CAMPS, built over TensorWeaver. The Clifft runs are performed with Clifft 0.7.0 with out-of-the-box parameters. Note that the peak bond dimensions $\chi_{peak}$ reported here for MPS are not proven to be minimal, and are instead the minimal found over our search of 1D layout. Note that all MPS simulations, both pure MPS and CAMPS, are performed in the exact regime without any truncations, up to machine precision.

As expected for MPS, its runtime depends mainly on $\chi_{peak}$, and is insensitive to the number of T gates present. Indeed, both \RegThFi schemes have similar runtime for the $\ket{T}$ or $\ket{S}$ version.  As also expected, the inverse is true for both CAMPS and Clifft, as the $\ket{S}$ version exhibits orders-of-magnitude faster runtime compared to the $\ket{T}$ cultivation.

Two things stand out from the runtimes. First, on the smaller distance circuits (\RegTh, \RotTh, \RotThFo), MPS and CAMPS give comparable performance for the $\ket{T}$ cultivation. While on larger circuits (\RegFi, \RotFi), CAMPS is 1 to 2 orders of magnitude faster than MPS. This can be explained by the disentangler of CAMPS managing to keep the peak bond dimension of the magic residual to 5 for both \RegFi and \RotFi, compared to 96 and 32 respectively for pure MPS, while it buys way less for the smallest circuits and still pays the disentangler work and the Clifford frame arithmetic. Second, Clifft is 2 to 3 orders of magnitude faster than CAMPS on all $\ket{T}$ circuits, except the \RegFi circuit. Indeed, in this circuit, the peak active dimension of Clifft --- the quantity in which its runtime is exponential in --- reaches 22, while CAMPS still keeps the peak bond dimension of the magic residual at 5. This results in CAMPS being $\sim13\times$ faster than Clifft. Additionally, the pure MPS engine reaches similar performance to Clifft. This is impressive as MPS is a general-purpose simulation framework, not made with near-Clifford circuits in mind.

Overall, this benchmark suggests that the \RegFi circuits represent a special class, where its magic part is present mainly in quantity ($k_{peak}=22$) but not in entanglement ($\chi_{peak, CAMPS}=5$), having overall low entanglement to begin with ($\chi_{peak, MPS}=96$). This shows that both CAMPS and frameworks such as the one Clifft uses have their place in the QEC simulation ecosystem.
\begin{table*}
    \centering
    \begin{tabular}{|c||c|c|c|c||c|c|c||}
    \cline{6-8}
        \multicolumn{5}{c ||}{} & \multicolumn{3}{c ||}{Runtimes} \\ 
        \hline
         &  \#qubits & \#gates & \#T & M & MPS $(\chi_{peak})$ & CAMPS $(\chi_{peak})$ & Clifft $(k_{peak})$\\
         \hline
        \RegTh           & 20 & 151 & 57 & 243 & $15.9~ms~(16)$    & $11.1~ms~(5)$       & $30.7~\mu s~(8)$ \\
        \RegTh$~\ket{S}$ & 20 & 139 & 0  & 233 & $15.2~ms~(16)$    & $80.0~\mu s~(1)$   & $17.9~\mu s~(0)$ \\
        \hline
        \RotTh           & 20 & 114 & 34 & 203 & $5.9~ms~(8)$      & $6.3~ms~(4)$       & $22.2~\mu s~(5)$ \\
        \textbf{Rot$_{3\times4}$}       & 23 & 147 & 77 & 303  & $13.2~ms~(16)$    & $14.7~ms~(5)$       & $31.2~\mu s~(8)$ \\
        \hline
        \RegFi           & 55 & 751 & 245& 1863 & $\bm{4.1~s~(96)}$ & $\bm{90.5~ms~(5)}$ & $\bm{1.2~s~(22)}$ \\
        \RegFi$~\ket{S}$ & 55 & 719 & 0  & 1847 & $2.4~s~(64)$      & $454.1~\mu s~(1)$   & $11.4~\mu s~(0)$ \\
        \hline
        \RotFi           & 55 & 599 & 163& 1584 & $609.6~ms~(32)$   & $47.3~ms~(5)$       & $173.8~\mu s~(12)$ \\
        \hline
    \end{tabular}
    \caption{Characteristics and run time for the circuits introduced in the main text using the qubit layout presented in \Cref{fig:layout_d3}, \ref{fig:layout_d5}, and \ref{fig:layout_newd5}. The number of T gates is counted after unitary decomposition of all non-Clifford gates ($CS$ and $CCZ$) in a T+Clifford basis. M is the number of depolarising channels in the circuits. Peak $\chi$ and runtime are given for the circuit with depolarising noise channels but with $p$ set to $0$. Runtime is the best over 10 runs of 1 shot, all engines are restricted to 1 thread, on an AMD Ryzen$^\text{TM}$ 7 CPU. The MPS simulations were performed on TensorWeaver 0.14.0.  Clifft runs are made with version 0.7.0 and default optimisation passes, and only include the "sample" call; pre-compilation is excluded. $\chi_{peak}$ is the maximum bond dimension observed on any bonds of the MPS (the full state for MPS, the magic residual CAMPS) after application of each operator. $k_{peak}$ is the maximum active dimension reported by Clifft.}
    \label{tab:runtime_noiseless}
\end{table*}

We used CAMPS and all the simulator-agnostic methods described above (subset-sampling, offline screening and online screening) to produce the results presented in \Cref{fig:results}, among which we present exact simulation of the Sahay \emph{et al.} circuit at d=5. For this circuit, we reach a sampling rate of $50~kHz$ with 16 processes parallelised over 16 threads, when sampling from the subsets with fault weights $w\geq5$ and following the Binomial distribution at $p=0.001$, as well as including the processing time required for the offline screening. We measured the equivalent 16-thread sampling rate of Clifft as $50.9 Hz$, using the native online screening and subset sampling\footnote{For Benchmarking purposes, and as Clifft only allows sampling from one fault weight subset at a time, we sampled independently from the subsets with fault weight from 5 up to 10 with a number of shots proportional to the corresponding factor $Bin_w$. We used the postselection mask option and the "survivor sampling" to implement the online screening, similar to the one used for the CAMPS simulation.}. This difference is in part due to the better performance of CAMPS for this circuit, but also due to the offline screening, which flags $99.8\%$ ($99.7\%$ for k=5) of the sampled shots as discarded, leaving $0.2\%$ to simulate with CAMPS.

\subsection{$\ket{T}$ vs $\ket{S}$ cultivation}
\label{sec:TvsS}
Being pure Clifford circuits and thus efficiently simulable, the $\ket{S}$ cultivation performances are usually used as a proxy for the $\ket{T}$ cultivation. However, it has been demonstrated \cite{gidneyMagicStateCultivation2024, liSOFTHighperformanceSimulator2025, chaseClifftFastExact2026} that the $\ket{S}$ and $\ket{T}$ versions of Gidney \emph{et al.} cultivation exhibit a discrepancy in logical error rate, with the ratio $T/S$ reaching up to $30$ on the d=5, depending on the noise strength. We show here that this is not the case for the fold-transversal protocol. For the \RegFi circuit our simulations give $T/S = 0.72$ with a $95\%$ credible interval $[0.41:1.21]$ at p=0.001, $0.80\ [0.45:1.37]$ at p=0.002, and $0.82\ [0.45:1.74]$ at p=0.003; the construction of the intervals is described in Appendix \ref{sec:credible_interval}. Note that the three intervals are correlated since they are constructed from the same samples for the $\ket{T}$ cultivation, reweighted through subset sampling. The ratio is slightly below unity throughout, and it shows no signs, within the confidence interval, of the order-of-magnitude difference reported for the d=5 Gidney \emph{et al.} protocol. Thus the $\ket{T}$ fold-transversal cultivation does not underperform its Clifford proxy, and might even have slightly better performance. Moreover, this shows that the $T/S$ ratio is not only distance-dependent, but also cultivation method dependent.

\subsection{Comparison of the schemes}
\label{sec:comparison}

In this section, we compare the performance of the different cultivation schemes described in \Cref{sec:MSC} and shown in Fig.~\ref{fig:results}. The two new cultivation schemes that we propose, \RotTh and \RotFi, are based on a partially fault-tolerant fold-transversal S gate in the rotated surface code. The partial fault tolerance means that those circuits can already show logical errors at fault weight 2 and 4, respectively, as opposed to 3 and 5 for the \RegThFi circuits.
This carries over to the logical error rate, which is, at $p=0.001$, $4$ to $6$ times higher, reaching $7.5\times10^{-7}$ and $8.1\times10^{-10}$ for d=3 and 5, respectively, while the regular surface code protocol reaches $1.8\times10^{-7}$ and $1.3\times10^{-10}$.

However, taking into account the survival probability of the shots and the lower gate count, we find that the \RotThFi schemes have a $\sim 1.3\times$ lower space-time overhead than their \RegThFi counterpart. The comparison is performed in more detail in Appendix \ref{sec:st_overhead}. Thus, they could be an interesting cheaper alternative whenever the better logical error rate of the \RegThFi is not necessary, especially considering that the escape stage should lower the discrepancy \cite{chaseClifftFastExact2026}, and additional noise, such as losses discussed in Appendix \ref{app:losses}, can increase the discrepancy between the schemes. The exact effect of the escape stage is left for future work

We also proposed the \RotThFo to try to mitigate the partial fault tolerance of the rotated surface code S gate by using an asymmetric rotated surface of distance 3 and 4 in the X and Z directions, respectively. We see that the logical error rate and the expected attempts are indeed brought back to the same order as the \RegTh scheme, showing, however, slightly worse performance on both counts. Thus we do not expect this scheme to be of any interest besides the fact that it only uses the rotated version of the surface code.

\section{Conclusion}


We have shown that the fold-transversal
surface-code cultivation circuit of Sahay \emph{et al.} produces weakly-entangled states, admitting compact untruncated MPS representations with noiseless peak
bond dimensions $\chi_{\text{peak}} = 16$ and $96$ at $d=3$ and $d=5$.
Adding Clifford frames lowers this further: the CAMPS disentangler holds the magic residual at $\chi_{\text{peak}} = 5$ for both
$d=5$ schemes, making CAMPS one to two orders of magnitude faster than pure MPS
there, and $\sim13\times$ faster than Clifft on \RegFi, whose peak
active dimension $k_{\text{peak}} = 22$ becomes expensive for the state-vector representation. This shows that magic that is large in quantity need not be large in entanglement, and near-Clifford simulators such as CAMPS and Clifft are
complementary.

Orthogonally, we introduced an offline screening step that decides by
stabiliser propagation alone the outcome of most measurements -- $91$ of the $133$ measurements of
\RegFi -- and paired with subsset sampling, discards, at $\mu$s per trajectory, up to $99.8\%$ of the
shots that post-selection would reject anyway. 
Combined with CAMPS, online screening and subset sampling, it brings the $d=5$ fold-transversal
circuit within reach. With this, we show that the $\ket{T}$ and $\ket{S}$ variants of the
fold-transversal scheme reach comparable logical error rates, with the credible
interval on $T/S$ consistent with or below unity across the noise range
considered. The order-of-magnitude gap reported for the colour-code scheme of
Gidney \emph{et al.} is therefore a property of the protocol and not of
cultivation as such, and the Clifford proxy cannot be assumed conservative
without checking.

We further proposed two cultivation schemes built on the partially
fault-tolerant transversal $S$ gate of the rotated surface code, and evaluated
them within the same framework. \RotTh and \RotFi reach
$\sim 1.3\times$ lower expected space-time cost than their
$\textbf{\sffamily Reg}$ counterparts, at the price of a logical error rate $\sim4-6$ times higher.

Several extensions follow naturally. The escape stage can be combined with our results through a simulator-agnostic hand-off simulation.
Separately, pure MPS allows access to a regime where Pauli-based simulators struggle: because the
representation is indifferent to magic, non-Clifford noise can be simulated directly, possibly at the cost of increasing entanglement.
For practical uses of MPS, automatic methods to optimize qubit ordering would be useful to develop, as by-hand optimization is hard in circuits with less obvious structure.
While the CAMPS disentangler manages to find decent frames to reduce entanglement, our ordering optimization shows that improvements are possible.
Finally, combining the active dimension reduction of Clifft with the MPS representation and disentangling procedure of CAMPS might lead to further improvements in simulation capability.

\paragraph*{Note added.} While this manuscript was being
finalised, we became aware of the related work of Takada
\textit{et al.}~\cite{takadaExactEfficientSimulation2026}, which introduces
Clifford-stabiliser simulation and applies it to the same
fold-transversal surface-code cultivation, reaching fault
distance 7. The two works are independent.

\section{Acknowledgment}
We thank Gerald E. Fux, Gavin K. Brennen and the QPerfect research and development team for inspiring discussions.
This work was supported by the French government through a 
CIFRE grant (Convention Industrielle de Formation par la Recherche) managed by the 
Association Nationale de la Recherche et de la Technologie (ANRT), convention No. 
2025/1625. The authors would like to acknowledge the High Performance Computing Center of the University of Strasbourg for supporting this work by providing scientific support and access to computing resources. Part of the computing resources were funded by the Equipex Equip@Meso project (Programme Investissements d'Avenir) and the CPER Alsacalcul/Big Data.

The CAMPS and MPS engines used here are part of QPerfect's MIMIQ platform and are not publicly available; MIMIQ is accessible commercially. 

\bibliography{references}
\bibliographystyle{quantum}

\appendix
\section{Details on simulation}
In this appendix, we describe the details of the simulations performed in this work.

\subsection{Circuits}
\label{app:circuit_description}

 Unless stated otherwise, all circuit primitives used here are described in detail in \cite{sahayFoldtransversalSurfaceCode2025}. 

For the $\ket{T}$ \RegThFi cultivation, we begin with the optimised unitary encoding and verification of the $\ket{T}$ state in the distance 3 rotated surface code as introduced by Sahay \emph{et al.}. The code is then brought to a regular surface code using half a QEC cycle. There, two $H_{XY}$ projective measurements are performed, with a GHZ state of 3 qubits. 

The $d=3$ cultivation stops here. To evaluate the performance of the circuit, we append an unphysical noiseless decoding circuit to decode the regular surface code, collapse and extract the values of the stabilisers with measurements, and evaluate the X,Y,Z expectation values of the state of the decoded qubits, equal to the logical state of the logical qubit before decoding. The decoding circuit is generated with the Munich Quantum Toolkit~\cite{mqt}. As the decoding circuit requires no additional qubits and consists solely of disentangling operations, it is an attractive option for simulation tools, such as MPS, whose cost grows with the number of qubits and/or entanglement, compared to a standard noiseless syndrome-extraction round in the final regular surface code.

After the first $H_{XY}$ checks, the $d=5$ cultivation continues with a unitary growth to a distance 5 rotated surface code with half a QEC cycle of the regular $d=3$ surface code. There, 3 full QEC cycles of the rotated surface code are performed. The rotated surface code is then morphed into a $d=5$ regular surface code, with half a QEC cycle of the rotated surface code. Two $H_{XY}$ projective measurements follow. The circuit ends with an unphysical noiseless decoding and syndrome/logical state extraction as for the distance 3 circuit.

For the $\ket{S}$ cultivation, we simply replace the encoded state in the unitary encoding with a $\ket{S}$ state, and the $H_{XY}$ checks with $Y$ checks.

\begin{figure*}
    \centering
    \includegraphics[width=0.7\textwidth]{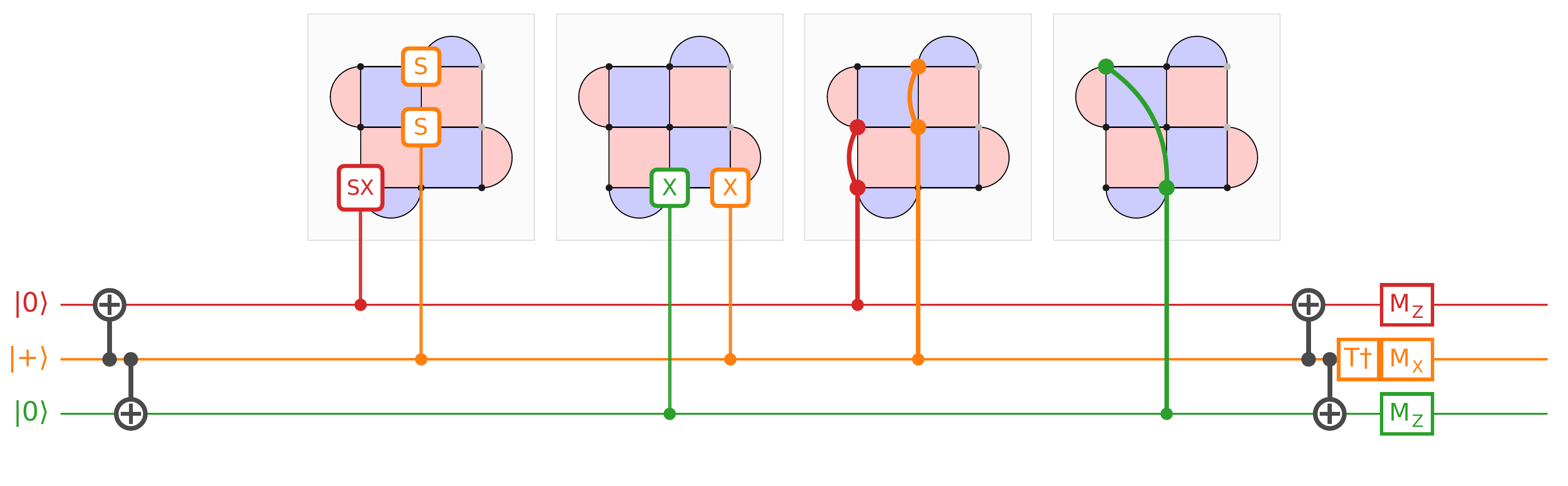}
    \caption{\textbf{Logical $\bm{ H_{XY}}$ check for Rot$\bm{_3}$.} The ancilla GHZ(3) state is supported on the bottom wires. The check is decomposed into controlled-S gates involving the diagonal qubits, controlled-X gates on the first row of qubits, and CCZ gates involving pairs of twisted off-diagonal qubits. 
    }\label{fig:hxychecksd3}
\end{figure*}

\begin{figure*}
    \centering
    \includegraphics[width=\textwidth]{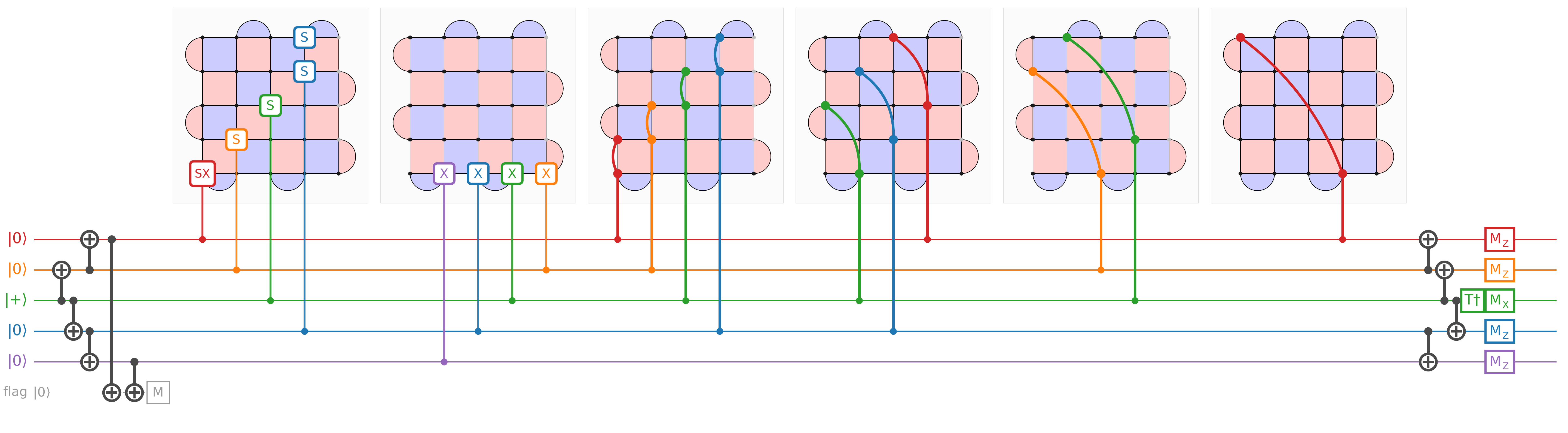}
    \caption{\textbf{Logical $\bm{ H_{XY}}$ check for Rot$\bm{_5}$.} The ancilla GHZ(5) state is supported on the bottom wires, with a 6th qubit serving as a flag for the state preparation. The check is decomposed into controlled-S gates involving the diagonal qubits, controlled-X gates on the first row of qubits, and CCZ gates involving pairs of twisted off-diagonal qubits. 
    }\label{fig:hxychecksd5}
\end{figure*}

The \RotThFi differs in several ways. First, the morphings between the rotated surface code and the regular surface code are not performed, and the target code will always stay in the rotated picture. The unitary growth is then replaced by a growth from the distance 3 to the distance 5 rotated surface code, following the circuit introduced in \cite{claesCultivatingStatesSurface2025} Figure 4. Additionally, the $H_{XY}$ checks are replaced with the rotated surface code variant, described in detail in \cref{fig:hxychecksd3} and \ref{fig:hxychecksd5}. The gate order and their GHZ qubit attribution are chosen such that no single (pair of) error(s) on the GHZ qubits can be propagated to a logical error on the d=3 (d=5) rotated surface code. The first layer of CCZ gates aligns with the logical Z operators, reducing by one the fault detection distance of the rotated surface code in the Z direction, and allowing a logical error of weight 2 and weight 4 for the d=3 and d=5 circuits, respectively. Finally, the unphysical noiseless unitary decoding is performed in the rotated surface code. The remaining of the circuits (unitary encoding, optimised verification and QEC cycles) are performed identically to the \RegThFi circuits.

\begin{figure*}
    \centering
    \includegraphics[width=\textwidth]{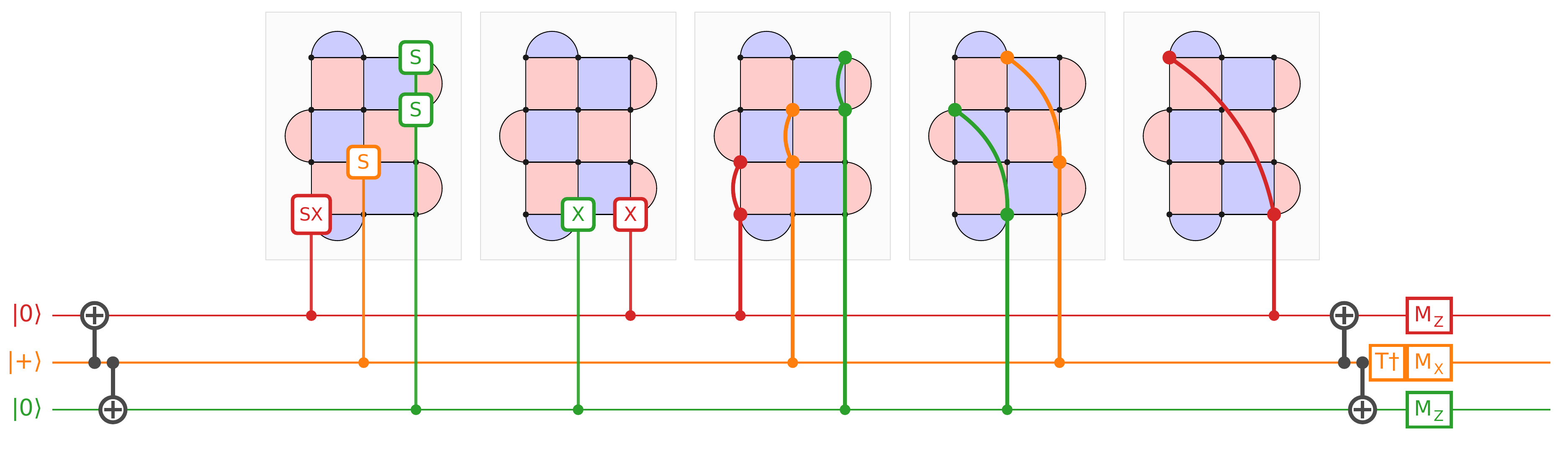}
    \caption{\textbf{Logical $\bm{ H_{XY}}$ check for Rot$\bm{_{3\times4}}$.} The ancilla GHZ(3) state is supported on the bottom wires. The check is decomposed into controlled-S gates involving the diagonal qubits, controlled-X gates on the first row of qubits, and CCZ gates involving pairs of twisted off-diagonal qubits. 
    }\label{fig:hxychecksd4}
\end{figure*}

In the \RotThFo, we use a surface code patch of distances $d_x=3$ and $d_z=4$ in the X and Z directions, respectively. The unitary injection is performed using the standard unitary encoding on the $d=3$ rotated surface code, augmented by a small morphing circuit to increase the distance in the Z direction, and followed by a modified version of the optimised stabiliser measurement to catch additional error propagation from the morphing circuit. Then the $H_{XY}$ operator checks are performed using the circuit described in \cref{fig:hxychecksd4}.

\subsection{Noise model}

All the simulations were performed with the same noise model, following the SD6 noise model of Sahay \emph{et al.} All unitary gates are preceded by a standard single- or multi-qubit depolarising noise channel of rate $p$. Measurements and resets in the Z basis are respectively followed and preceded by a Pauli X error channel of probability $p$. Idle noise is standard depolarising channels of probability $p$ applied on idle qubits. For all the $H_{XY}$ and $Y$ checks, the idle qubits are determined for a compact layout, where we use independent GHZ states prepared and measured in parallel with the instructions performed on the main surface code patch.

\subsection{Post-selection and logical error}

All measurements, including the ones from the syndrome extraction of the final noiseless unitary decoding, are used as standalone detectors. This means that the shot is discarded if any of the measurements returns 1.

A logical error is flagged as soon as the squared overlap between the output logical state (collapsed to a single qubit by the noiseless decoding circuit) and the target state ($\ket{T}$ or $\ket{S}$) is not exactly $1$ up to machine precision. Note that since the noise model used in this work only contains Paulis, the overlap will take significantly different values from 1 whenever a logical error occurs.

\subsection{MPS qubits ordering}
\label{app:qubit_layout}

\begin{figure}
    \centering
    \includegraphics[width=\linewidth]{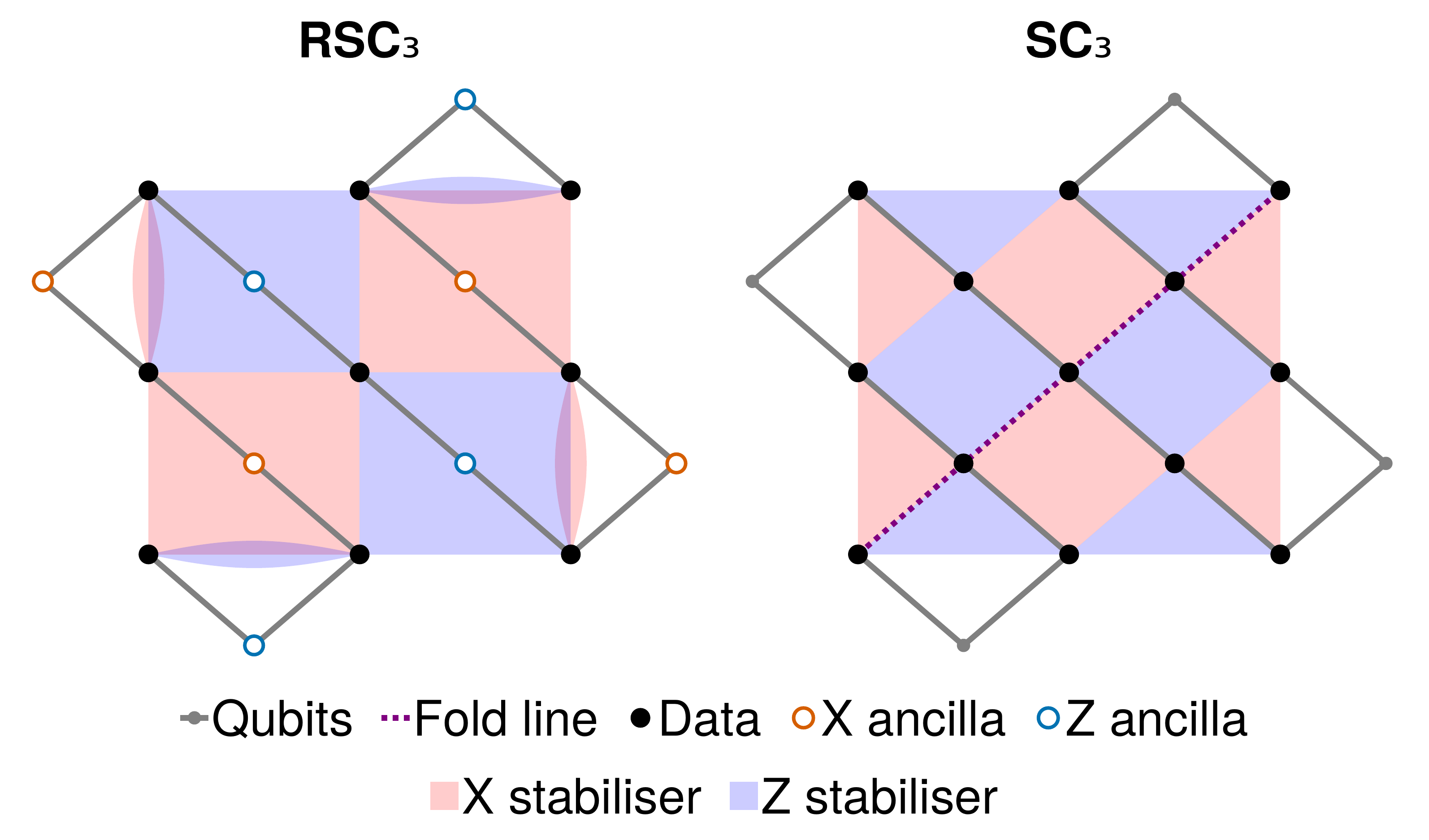}
    \caption{\textbf{Qubit layout for the \RegTh and \RotTh cultivation MPS simulation.} The distance 3 rotated surface code (RSC$_3$) is used at the injection, while the distance 3 regular surface code (SC$_3$) is used during the $H_{XY}$ operator checks for \RegTh. The \RotTh scheme only use the RSC$_3$.}
    \label{fig:layout_d3}
\end{figure}

\begin{figure}
    \centering
    \includegraphics[width=\linewidth]{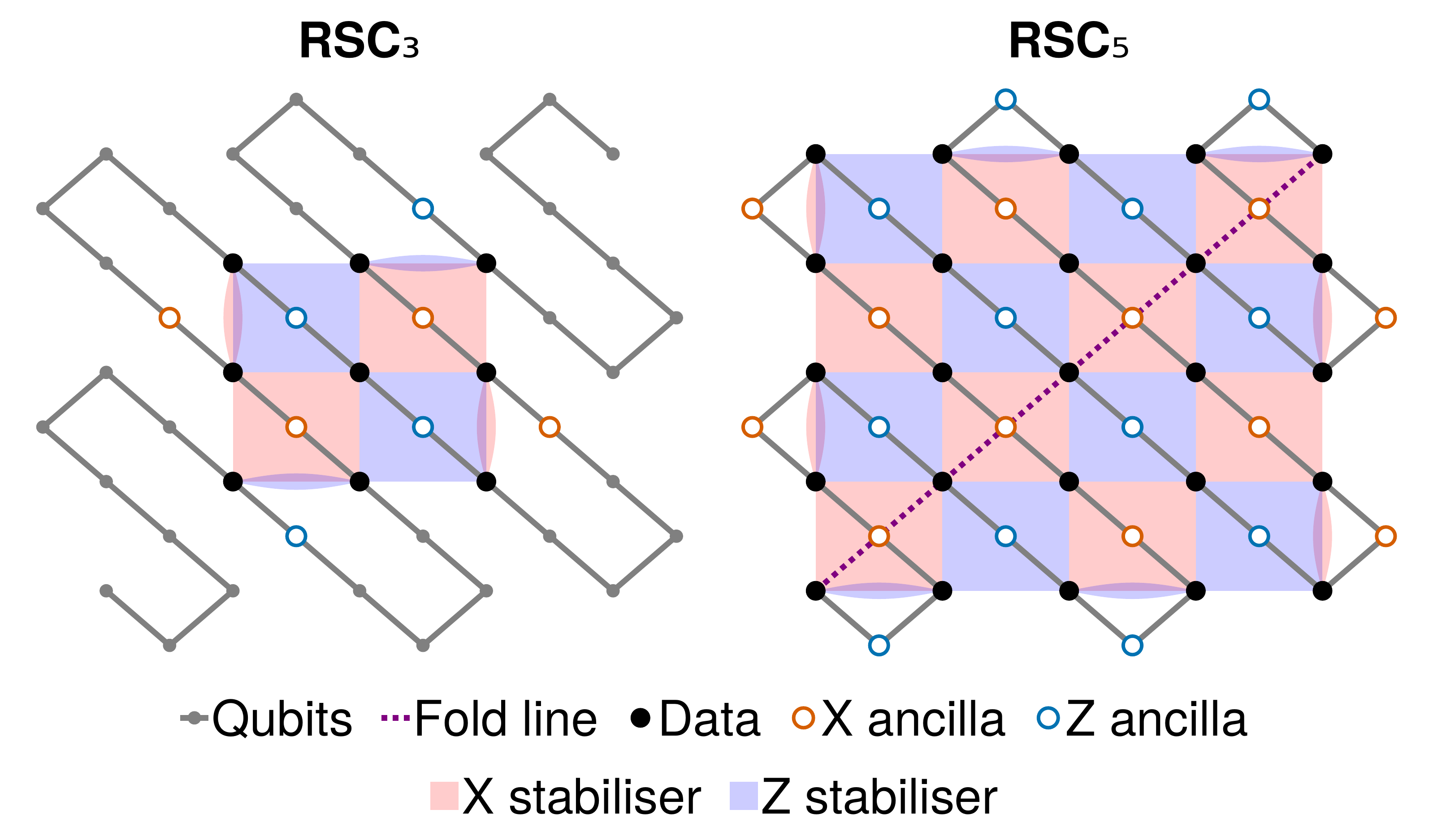}
    \caption{\textbf{Qubit layout for the \RotFi cultivation MPS simulation.} The distance 3 rotated surface code (RSC$_3$) is used at the injection stage, and the first two checks, then the distance 5 rotated surface code (RSC$_5$) is used for QEC cycles and the last two checks.}
    \label{fig:layout_newd5}
\end{figure}

\Cref{fig:layout_d3} and \ref{fig:layout_newd5} describe the qubit layout used for MPS for, respectively, the \RegTh and \RotTh circuits, and the \RotFi circuit. The GHZ state qubits and the corresponding flag always sit at the end of the chain, since placing them inside the chain, closer to the qubits they act on during the checks, always resulted in higher bond dimension and thus higher runtime.

\section{Space-time overhead}
\label{sec:st_overhead}

\begin{figure*}
    \centering
    \includegraphics[width=\linewidth]{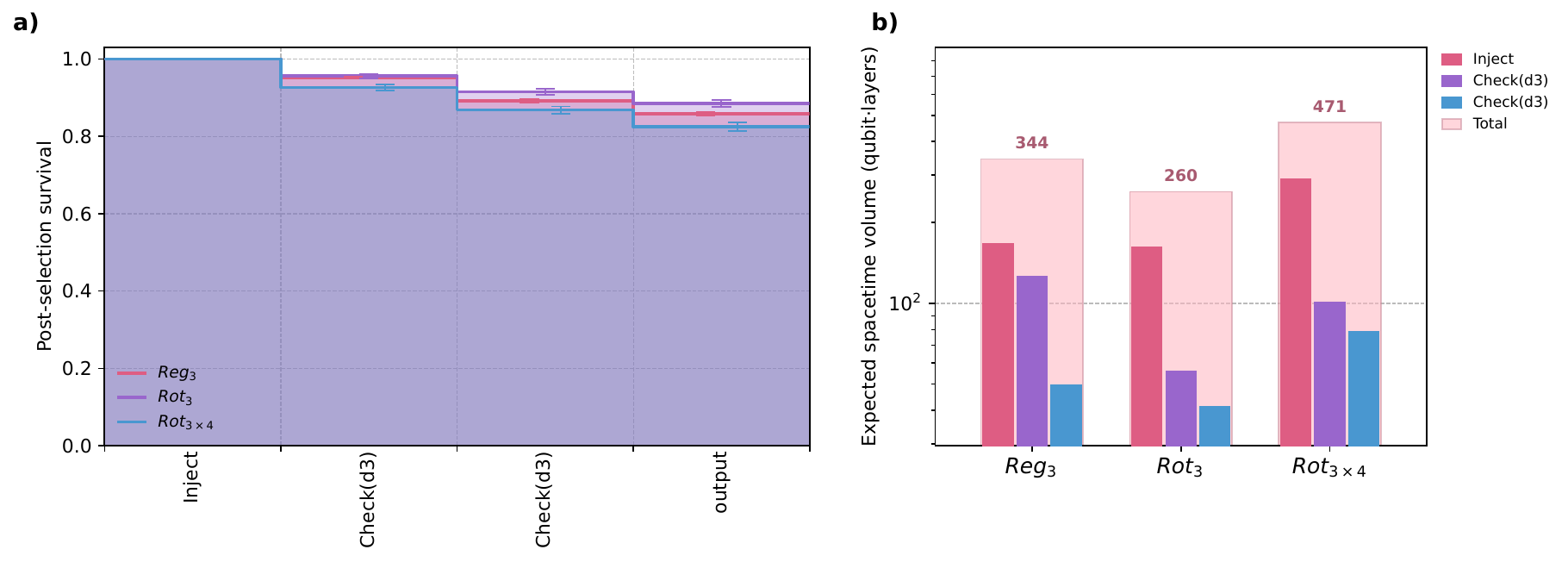}
    \caption{\textbf{Space-time overhead breakdown for the d=3 $\ket{T}$ cultivation circuits, excluding escape stage, at p=0.001.} 
    (a) Survival rate histogram of the shots at each post-selection step of the circuits. The first check stage includes the morphing from the rotated to the regular surface code for the \RegTh circuit. 
    (b) Comparison of the total expected space-time volume of the schemes, with a breakdown of each section's contribution.
    Note that the noiseless syndrome extraction and the post-selection based on its measurement are not included, as not physical. \errobars}
    \label{fig:st_d3}
\end{figure*}

\begin{figure*}
    \centering
    \includegraphics[width=\linewidth]{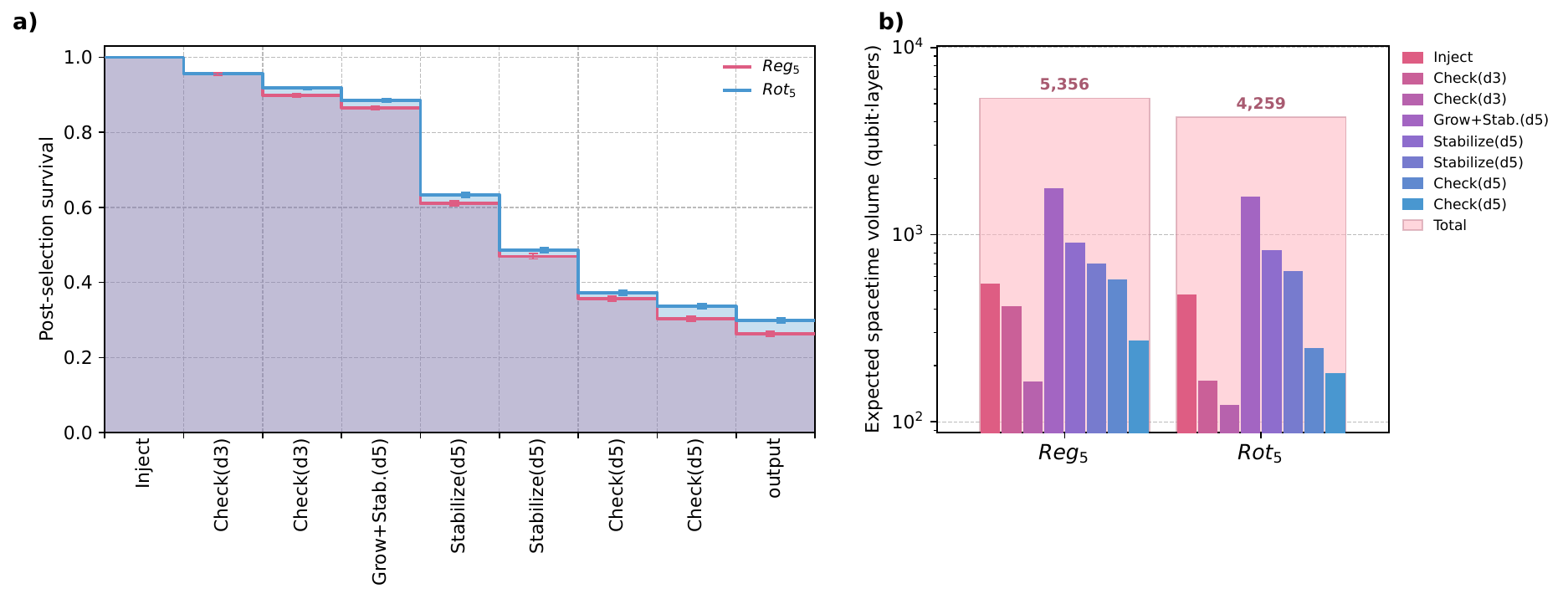}
    \caption{\textbf{Space-time overhead breakdown for the d=5 $\ket{T}$ cultivation circuits, excluding escape stage, at p=0.001.} 
    (a) Survival rate histogram of the shots at each post-selection step of the circuits. The first distance 3 and 5 checks stages include the morphing from the rotated to the regular surface code for the \RegFi circuit. 
    (b) Comparison of the total expected space-time volume of the schemes, with a breakdown of each section's contribution.
    Note that the noiseless syndrome extraction and the post-selection based on its measurement are not included, as not physical. \errobars}
    \label{fig:st_d5}
\end{figure*}

We computed the expected space-time overheads of all the proposed schemes for $\ket{T}$ state preparation using the methodology described in \cite{sahayFoldtransversalSurfaceCode2025} Appendix E.3. The space-time volume of a section of circuit is counted as the product of active (used) qubits and the number of gate layers (i.e. the circuit depth). Then the total expected volume is accounted for by summing the volume of each section of the circuit, weighted by the actual fraction of shots that reach that part of the circuit before post-selection. This assumes that the execution of the circuit is stopped as soon as the accumulated measurements allow deciding that the shot will be discarded. The survival rates across all the circuit segments are reported in \cref{fig:st_d3} and \ref{fig:st_d5} panel (a), while the corresponding panels (b) compare the expected space-time volume of the different schemes and provide a breakdown of the contribution of the different circuit segments. Note that, contrary to the results in \cref{fig:results}, the noiseless syndrome extraction and the post-selection based on its measurement are not included, as not physical.

\section{Losses}
\label{app:losses}
Neutral atom and ion quantum computers are subject to qubit losses, where an atom or ion is lost during an operation. Using the post-selection of the cultivation scheme, we show here that the impact of losses can be analytically estimated.

We suppose that all losses during the injection/cultivation phase can be perfectly detected, either through the already present measurements or with a loss detection unit (LDU) \cite{perrinQuantumErrorCorrection2025} appended at the end of the circuit. Note that the LDU would add layers of noisy gates, which we choose not to include in the analysis. However, this layer of noisy gates could be approximated by simply adding a layer of noise channels at the end of the circuit. Additionally, we can reasonably suppose that losses are modelled using independent loss channels, uncorrelated with the depolarising channels, and that we choose to discard each shot with a loss. In this regime, losses will only contribute to the discard rate, and not to the Logical error rate as:
$$
p_{acc}= (1-p_L)(1-p_D)
$$
where $p_{acc}$, $p_L$, and $p_D$ are respectively the probabilities for a shot of being conserved, having a loss, and having a flipped detector in the lossless case. The value of $p_D$ is given by the lossless simulations, while the value of $p_L$ can be exactly calculated given the loss model. 
Then the impact on the retry rate $1/p_{acc}$ is the multiplication by a factor $1/(1-p_L)$ compared to the lossless case. For example, for M independent loss channels, all with probability $p_l$ of loss, then $p_L=1-(1-p_l)^M$ and the multiplication factor is exactly $(1-p_l)^{-M}$. Combined with subset-sampling, this allows to reuses the same samples to probe an entire grid of $p$ and $p_l$ values, at no additional costs.

\begin{table}[]
\centering
    \begin{tabular}{|l | c | c | c| c|}
    \hline
    $p_{loss}$      & 0.0001  & 0.0005 & 0.001  & 0.002    \\
    \hline
    \RegTh  & 1.02  & 1.13  & 1.28  & 1.63 \\
    \RegTh$\ket{S}$  & 1.02  & 1.12  & 1.26  & 1.59 \\
    \hline
    \RotTh  & 1.02  & 1.11  & 1.23  & 1.5 \\
    \RotThFo  & 1.03  & 1.16  & 1.35  & 1.83 \\
    \hline
    \RegFi  & 1.2  & 2.54  & 6.45  & 41.67 \\
    \RegFi$\ket{S}$  & 1.2  & 2.52  & 6.35  & 40.35 \\
    \hline
    \RotFi  & 1.17  & 2.21  & 4.88  & 23.84 \\
    \hline
    \end{tabular}
    \caption{\textbf{Multiplicative coefficient on the expected number of attempts due to losses}, on the various circuits explored in this work, and at four different levels of loss rate.}\label{tab:loss}
\end{table}

\begin{figure}
    \centering
    \includegraphics[width=\linewidth]{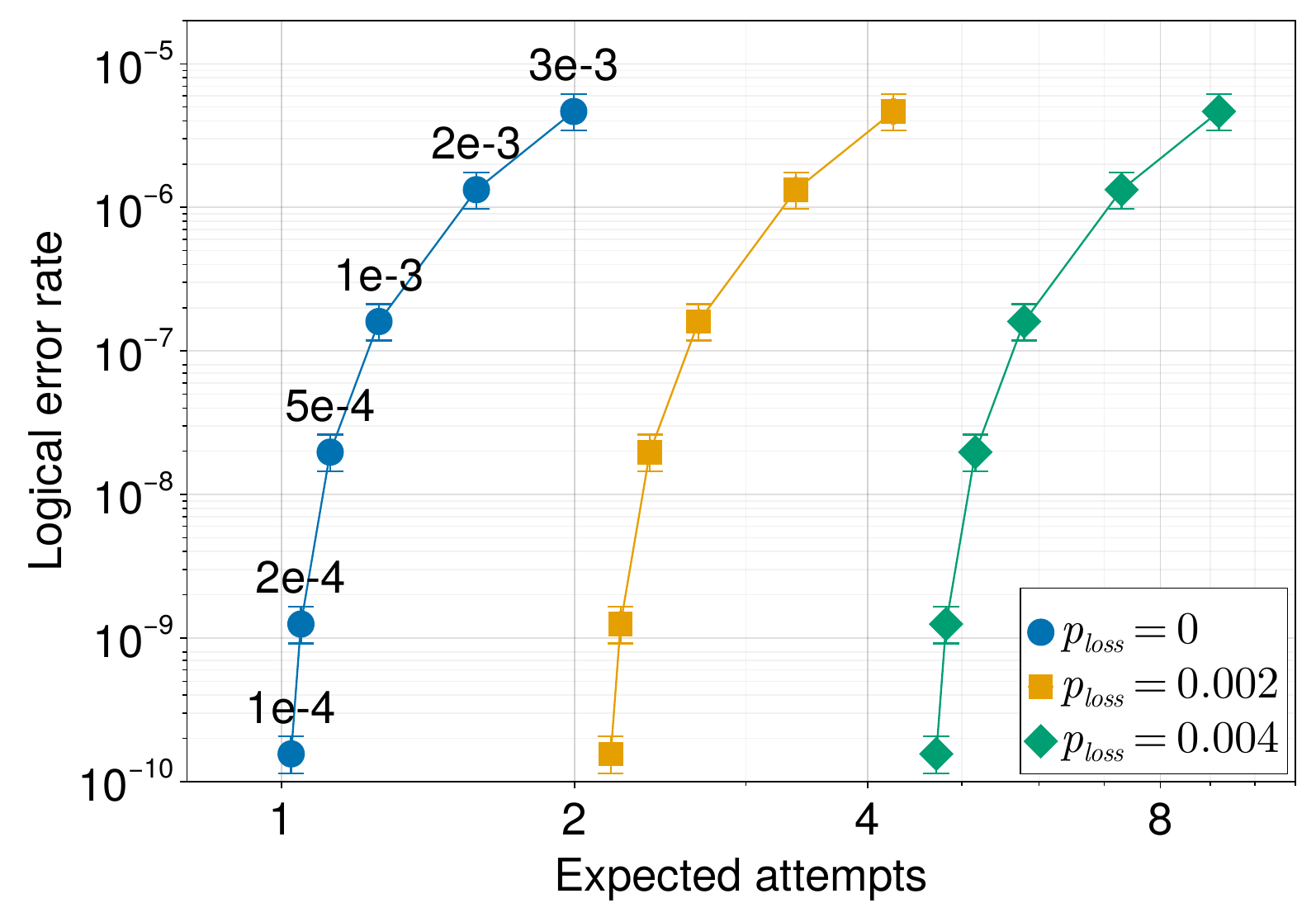}
    \caption{\textbf{Performance of the \RegTh scheme under losses.} The loss probability $p_{loss}$ is given in the legend, while the depolarising probability is given by the point label. \errobars}
    \label{fig:losses}
\end{figure}

Following those assumptions, and assuming that each depolarising channel in the circuits carries an independent loss channel of probability $p_{loss}$, we evaluated the impact of losses on the expected number of attempts of all the previously introduced schemes, recorded as the multiplicative factor on this quantity due to loss alone. We compiled the values obtained in \cref{tab:loss}. We see that losses can become costly in the bigger circuits. This opens the possibility of discarding only some loss configurations to mitigate this effect. We leave this for future work.
To illustrate that losses only act on the expected attempts and not on the logical error rate, \Cref{fig:losses} shows the resulting performance for various noise models, including depolarising and loss channels, respecting the hypothesis stated above.

\section{Error analysis of the subset-sampling estimator}
\label{app:error-analysis}

This appendix gives the interval construction behind the error bars quoted on
$p_{acc}$ and $p_{le}$, and the three systematics that sit outside those intervals.

\subsection*{Per-bin likelihood}

Bin $w$ contributes $n_w$ shots, each with three mutually exclusive outcomes ---
rejected by post-selection, accepted and correct, accepted and wrong --- of
probabilities $1-p^{(w)}_{acc}$, $p^{(w)}_{acc}(1-p^{(w)}_{le})$ and
$p^{(w)}_{acc}p^{(w)}_{le}$. The counts $(n_w-n_w^{acc},\,n_w^{acc}-n_w^{err},\,
n_w^{err})$ are therefore multinomial, and the likelihood --- the probability of
observing those counts, read as a function of the two unknown rates --- is their
product, up to a combinatorial prefactor that carries no parameter dependence and
cancels in every ratio below. Collecting the powers of $p^{(w)}_{acc}$, both
accepted outcomes carrying one, the log-likelihood splits into two independent
binomial terms,
\begin{equation}\label{eq:loglik-split}
\begin{split}
    \ell_w &= \underbrace{n_w^{acc}\log p^{(w)}_{acc}
        +\big(n_w-n_w^{acc}\big)\log\big(1-p^{(w)}_{acc}\big)}_{n_w\ \text{trials}} \\
    &\;+\;\underbrace{n_w^{err}\log p^{(w)}_{le}
        +\big(n_w^{acc}-n_w^{err}\big)\log\big(1-p^{(w)}_{le}\big)}
        _{n_w^{acc}\ \text{trials}} \raisetag{3\baselineskip} ,
\end{split}
\end{equation}
the first counting how many of the $n_w$ shots survive post-selection, the second how
many of those $n_w^{acc}$ survivors then fail. The split is exact rather than
asymptotic: it is the sequential decomposition of the multinomial into an acceptance
draw followed by a failure draw whose number of trials is the outcome of the first.

Since no term contains both parameters, the mixed second derivative of $\ell_w$
vanishes identically and the two are orthogonal. Three consequences follow. The
maximum-likelihood estimates decouple into Eq.~\eqref{eq:estimators};
$p^{(w)}_{le}$ is exactly a binomial proportion over accepted shots, with no random
denominator to correct for; and imposing a value on one parameter leaves the
estimate and the interval of the other untouched. The last point is what licenses
the analytically fixed bins of the main text, where $p^{(w)}_{le}=0$ is imposed for
$w<d_e$ while $p^{(w)}_{acc}$ continues to be estimated from data in those same
bins.

The information available on $p^{(w)}_{le}$ still
depends on $p^{(w)}_{acc}$, which controls how many trials the second term receives.
In particular a bin with $n_w^{acc}=0$ carries no information at all on
$p^{(w)}_{le}$, which becomes unidentified: the second term of
Eq.~\eqref{eq:loglik-split} vanishes for every value of the parameter. The bin is
not lost, since its $n_w$ shots still constrain $p^{(w)}_{acc}$. Under the
constrained maximisation below, the free $p^{(w)}_{le}$ is simply driven to the
endpoint favouring the constraint, so the bin adds at most
$\mathrm{Bin}_w(p)\,p^{(w),+}_{acc}$ to the upper limit on the numerator of
$p_{le}$ and nothing to the lower one. We verify that this worst case is negligible
rather than requiring an acceptance in every bin; since
$p^{(w),+}_{acc}\simeq\log R/n_w$, with $R$ the likelihood-ratio threshold
introduced below, it can be suppressed by sampling even when no shot in the bin is
ever accepted.
 
\subsection*{Intervals}
 
We use likelihood-ratio intervals throughout: for a single parameter, the interval
is the set of values whose likelihood lies within a factor $R$ of the maximum, and
we take $R=100$, corresponding asymptotically to $99.8\%$ coverage, i.e. roughly a
three-sigma statement. The quantities we report, however, are not individual bin
rates but two functionals of the whole set $\{p^{(w)}_{acc},p^{(w)}_{le}\}$: the
weighted sum $p_{acc}$ and the ratio of weighted sums $p_{le}$. We therefore
construct the interval for the functional itself rather than assembling it from
per-bin intervals. Collecting all bin rates into
$\vec{p}=\{p^{(w)}_{acc},p^{(w)}_{le}\}_w$ and writing
$\ell(\vec{p})=\sum_w\ell_w$ for the joint log-likelihood --- a sum because the
bins are sampled independently --- the interval reported on a functional
$\theta=f(\vec{p})$ is
\begin{equation}\label{eq:profile}
    \Big\{\,\theta\;:\;
    \max_{f(\vec{p})=\theta}\ell(\vec{p})
    \;\ge\;
    \max_{\vec{p}}\,\ell(\vec{p})-\log R\,\Big\}.
\end{equation}
In words: for a candidate value $\theta$, re-maximise the likelihood over all bin
rates subject to the functional taking that value, and retain $\theta$ if the cost
in log-likelihood stays below $\log R$. The endpoints are located by bisection on
that constrained maximum.
 
The constrained maximisation is also what handles the correlation between the
numerator and the denominator of $p_{le}$, which are computed from the same shots
and are not independent: here they are simply two functions of one common parameter
set, so no covariance has to be modelled. The cheaper alternative --- fitting each
bin separately, scaling its interval by $\mathrm{Bin}_w$, summing in quadrature and
propagating through the ratio by the delta method --- agrees with
Eq.~\eqref{eq:profile} where all counts are large, but quadrature presumes symmetric
errors and the per-bin intervals are strongly asymmetric at the counts we work with.
 
To avoid relying on large-number asymptotics, which the rare failure counts in the bins that matter do not necessarily support, we directly measured the coverage of Eq.~\eqref{eq:profile}. The procedure is a parametric bootstrap. We take the fitted per-bin
rates as though they were the truth; draw fresh counts from them, holding each $n_w$
at its realised value; rerun the same estimator on those counts; and record whether
the resulting interval still contains the value that assumed truth implies, computed
with the same truncation as the reported numbers. The fraction of replicates that
cover is then compared with the nominal $99.76\%$. Over five protocols, three target error rates, four assumed truths and $5000$
replicates each, the coverage
pooled to $99.760(10)\%$, and no single configuration fell below $99.56\%$. The
largest deviation among the configurations was $2.9$ standard errors, about
what the spread of that many should produce.

\subsection*{Systematics}

Three effects are not captured by Eq.~\eqref{eq:profile}.

\emph{Truncation.} Cutting the sums of Eq.~\eqref{eq:reweight} at $w_{max}$ omits
from each at most $B=\sum_{w>w_{max}}\mathrm{Bin}_w(p)$, using
$p^{(w)}_{acc},p^{(w)}_{le}\le1$. For $p_{acc}$ the omission is
purely one-sided and $B$ is added to the upper endpoint, rather than in quadrature.
For the ratio $p_{le}$ it is not quite one-sided, since the missing mass sits in
both numerator and denominator; but $w_{max}$ is chosen so that $B$ falls orders of
magnitude below the reported rate $p_{le}p_{acc}$, and the downward component is
then smaller than the upward one by a further factor of order $p_{le}$, so we
neglect it and again add $B$ upward only.

\emph{Unsampled bins.} A bin below $w_{max}$ that the sampling never reached,
$n_w=0$, is silently absent rather than bounded, which is the more dangerous
failure. We remove this by spending some sampling effort uniformly across the fault weight range 1 to $w_{max}$. This is distinct from a bin that was
sampled but never accepted, $n_w>0$ with $n_w^{acc}=0$: there the shots do bound
the bin's contribution, which is already carried inside
Eq.~\eqref{eq:profile}, and no separate systematic is needed.

\emph{The value of $d_e$.} Overestimating $d_e$ omits
$\sum_{w<d_e}\mathrm{Bin}_w\,p^{(w)}_{acc}p^{(w)}_{le}$ with nothing in the interval
to signal it, and the binomial weights involved are the large ones. Since the circuit
is not Clifford, $d_e$ cannot be obtained from a minimum-weight search over a
detector error model, and rests instead on a structural analysis of the cultivation
circuit.

\section{Credible intervals for rate ratios}
\label{sec:credible_interval}
The ratio $R=T/S$ of cultivation error rates is reported with a $95\%$ central credible interval obtained by combining the two posteriors directly. For the $\ket{S}$ cultivation, sampled with stim, the posterior on the post-selected logical error rate is exact: $S\sim Beta(n_{err} + 1/2,~n_{acc}-n_{err}+1/2)$, the Jeffreys posterior for a binomial rate, with $n{acc}$ the number of accepted shots and $n_{err}$ the number of those carrying a logical error. For the $\ket{T}$ cultivation, the posterior is taken as the normalised profile likelihood, $\Pi(\theta)\propto \exp(-[l_{max} - max_{f(p_0)=\theta}~l(p_0)])$, evaluated on a log-spaced grid of $\theta$ extended until the profile has dropped 18 log-units below its maximum on both sides (leaving <$10^{-8}$ of the mass outside the grid), with a prior flat in $\log(\theta)$; the same profile-likelihood function is used here as for the likelihood-ratio intervals of Appendix \ref{app:error-analysis}, so no separate calculation enters. The density of $\log(R) = \log(T)-\log(S)$ is then obtained by numerical convolution of the two log-densities on a shared lattice, and the interval is read off as its 2.5 and 97.5 percentiles. Treating the profile likelihood as a
posterior for $\theta$ alone is the only approximation in this construction.

\end{document}